\documentclass[11pt]{article}
\usepackage{graphics}
\usepackage[dvips]{graphicx}
\usepackage{amssymb}
\usepackage{amsmath}
\usepackage{vmargin}
\usepackage{feynmf}
\fmfcmd{%
style_def wiggly_arrow expr p =
cdraw(wiggly p);
shrink (2);
cfill (arrow p);
endshrink;
enddef;}

\setpapersize{A4}

\begin{document}
\setlength{\textwidth}{15cm}
\setlength{\textheight}{22cm}

\title{Lagrangian statistical mechanics applied to non-linear stochastic field
equations}
\author{Sam F. Edwards\\Cavendish Laboratory, University of Cambridge, UK\\and\\ Moshe Schwartz\\
School of Physics and Astronomy\\Raymond and Beverly Sackler
Faculty of Exact Sciences\\Tel Aviv University\\ Tel Aviv 69978,
Israel} \maketitle

\Large{\textbf{Abstract}}

We consider non-linear stochastic field equations such as the KPZ
equation for deposition and the noise driven Navier-Stokes
equation for hydrodynamics. We focus on the Fourier transform of
the time dependent two point field correlation,
$\Phi_{\bf{k}}(t)$. We employ a Lagrangian method aimed at
obtaining the distribution function of the possible histories of
the system in a way that fits naturally with our previous work on
the static distribution. Our main result is a non-linear
integro-differential equation for $\Phi_{\bf{k}}(t)$, which is
derived from a Peierls-Boltzmann type transport equation for its
Fourier transform in time $\Phi_{\bf{k}, \omega}$. That transport
equation is a natural extension of the steady state transport
equation, we previously derived for $\Phi_{\bf{k}}(0)$. We find a
new and remarkable result which applies to all the non-linear
systems studied here. The long time decay of $\Phi_{\bf{k}}(t)$ is
described by $\Phi_{\bf{k}}(t)~\sim ~
\exp\left(-a|{\bf k}|t^{\gamma}\right)$, where $a$ is a constant
and $\gamma$ is system dependent.

\section{Introduction}
\label{intro}
\setcounter{equation}{0}

Non-linear field equations describing a system with friction,
non-linearity and a driving noise have received much attention.
Two prime examples are the KPZ equation \cite{ka86} for the height
of a non-linearly deposited material,

\begin{equation}
\frac{\partial h}{\partial t} + \nu\nabla^2 h +\left(\nabla h\right)^2 = f
\label{11}
\end{equation}
and the Navier-Stokes equation
\begin{equation}
\frac{\partial {\bf u}}{\partial t} + \nu\nabla^2{\bf u} + {\bf u}\cdot\nabla{\bf u} -\frac{\nabla p}{\rho} = {\bf f}
\label{12}
\end{equation}
where $p$ is determined by $\nabla\cdot\bf{u}\rm{=}0$.

The noise, $f$, is considered Gaussian, stochastic driven, with
\begin{equation}
\left< f\left(\bf{r}\it{,}t\right) f\left(\bf{r}\it{'},t'\right)\right>=H\left( \bf{r} \rm{-} \bf{r}\it{'},t-t'\right)
\label{13}
\end{equation}
We write the equations in a general form in Fourier transform:
\begin{equation}
\frac{\partial h_{\bf{k}}}{\partial t}\,+\,
\nu_{\bf{k}}\,h_{\bf{k}} \,+\,
\sum_{j,l}\,M_{\bf{{\bf kjl}}}\,h_{\bf{j}}\,h_{\bf{l}} \,+\,
\sum_{\it{j,l,m}}N_{\bf{{\bf kjl}m}}\,h_{\bf{j}}\,h_{\bf{l}}\,h_{\bf{m}}
\,{\cdots}\, =\, f_{\bf{k}}\it{\left(t\right)} \, , \label{14}
\end{equation}
where we write $h$ as a scalar, but a minor elaboration of
notation covers vectors, \textit{e.g.}
$M_{\bf{{\bf kjl}}}^{\rm{\alpha\beta\gamma}}u_{\bf{j}}^{\rm{\beta}}u_{\bf{l}}^{\rm{\gamma}}$.
$M$ is taken independent of the origin, and therefore contains
$\delta_{\bf{k},\,\bf{j}\rm{+}\bf{l}}$. Note that we are using
here box normalization, namely the Fourier transform is defined as
$h_{\bf{k}}=\frac{1}{\sqrt{V}}\int h({\bf r}) e^{i\bf k\cdot
r}\,{\rm d^3} {\bf r}$, where $V$ is the volume of the system.
Consequently our $M$'s are order of $1/\sqrt{V}$ and the $N$'s are
order of $1/V$ etc. If (\ref{14}) is also Fourier transformed in
time:
\begin{equation}
i\omega
h_{\bf{k}\rm{\omega}}+\nu_{\bf{k}}\rm{h}_{\bf{k}\rm{\omega}} +
\sum_{j,l,\sigma ,\theta }
M_{\bf{k}\rm{\omega}\bf{j}\rm{\sigma}\bf{l}\rm{\theta}}h_{\bf{j}\rm{\sigma}}
h_{\bf{l}\rm{\theta}} + \cdots = f_{\bf{k}\rm{\omega}}\,,
\label{15}
\end{equation}
where $M$ now also contains $\delta_{\omega,\sigma + \theta}$. In
earlier papers, we approached the steady state solution of the
system (\ref{11}) by deriving the Liouville equation for the
probability $P\left( h,t \right)$,
\begin{equation}
\frac{\partial P}{\partial t} - \sum\frac{\rm{\partial}}{\partial h_{\bf{k}}}\it{\left( \nu_{\bf{k}} h_{\bf{k}} + \sum_{j,l} M_{\bf{{\bf kjl}}}h_{\bf{j}}h_{\bf{l}} - f\right)}P=0
\label{16}
\end{equation}
which when averaged over $f$, taking this to be the noise with
\begin{equation}
H=D\left( {\bf r}-{\bf r}'\right)\delta\left( t-t'\right)\rightarrow D_{0{\bf{k}}}\delta\left(t-t'\right),
\label{17}
\end{equation}
satisfies the well known form
\begin{equation}
\frac{\partial P}{\partial t} - \sum_{{\bf k}} \frac{\partial}{\partial h_{{\bf k}}}\left( D_{0{\bf k}}\frac{\partial}{\partial h_{-{\bf{k}}}} + \nu_{{\bf k}} h_{{\bf k}} + \sum_{{\bf j},{\bf l}} M_{{\bf k}{\bf j}\bf{ l}}h_{{\bf j}}h_{{\bf l}}\right)P=0
\label{18}
\end{equation}
$P$ is now the average over $f$ of equation (\ref{16}), and the steady state satisfies:
\begin{equation}
\sum\frac{\partial}{\partial h_{{\bf k}}}\left( D_{0{\bf k}}\frac{\partial}{\partial h_{-{\bf k}}} + \nu_{{\bf k}} h_{{\bf k}} +\sum_{{\bf j},{\bf l}} M_{{\bf k}{\bf j}{\bf l}}h_{{\bf j}}h_{{\bf l}}\right) P=0
\label{19}
\end{equation}

The approach to equation (\ref{19}) is to derive a transport equation based on a self consistent method, \textit{i.e.} suppose that the system can be developed about the model
\cite{SE92,SE98}:
\begin{equation}
\sum \frac{\partial}{\partial h_{{\bf k}}}\left( D_{{\bf k}}\frac{\partial}{\partial h_{-{\bf k}}} + \omega_{{\bf k}} h_{{\bf k}} \right) P_0 = 0
\label{110}
\end{equation}
\textit{i.e.}
\begin{equation}
P_0=\mathcal{N}\exp\left( -\frac{1}{2}\sum_{{\bf k}}\omega_{{\bf k}} \frac{h_{{\bf k}}h_{-{\bf k}}}{D_{{\bf k}}}\right)
\label{111}
\end{equation}
where
\begin{eqnarray}
\left< h_{\bf{k}}\it{h_{-{\bf k}}}\right> &  = & \phi_{\bf{k}} \\
\phi_{{\bf k}} = & \frac{D_{{\bf k}}}{\omega_{{\bf k}}},
\label{112}
\end{eqnarray}
$\phi_{{\bf k}}$ being the true two point function. In Peierls'
treatment of non-linear crystal electricity \cite{pei}, $\phi_{{\bf k}}$
appears as the number of phonons $n_{\bf k}$ and satisfies the
Boltzmann equation. In turbulence $\phi_{{\bf k}}$ is the energy
in the mode $\bf{k}$. In granular deposition there is no name for
$\phi_{{\bf k}}$ but perhaps we can call it the "flucton" since it
measures the surface fluctuation. Peierls could use perturbation
theory to derive the kernel of his Boltzmann equation, but since
the non-linearity dominates our problem we need both $\phi_{{\bf
k}}$ and $\omega_{\bf{k}}$. $P_0$ is the approximate solution of
equation (\ref{18}), which starts the self consistent expansion,
which is given in \cite{SE92,SE98} symbolically by:
\begin{multline}
P=P_0\bigg( 1-\sum \frac{Mhhh/\phi}{\omega + \omega + \omega} +\sum
\frac{MM(hhhh)/\phi\phi)hh}{\sum\omega\sum\omega}+\hbox{similar
terms}\\+\sum\frac{MMM(hhh/\phi)(hhh/\phi)(hhh/\phi)}{\sum\omega\sum\omega\sum\omega}+\hbox{similar
terms}+\cdots \bigg)
\label{113}
\end{multline}
pictures of theses terms are given in Appendix A below. It was
shown in refs \cite{SE92,SE98}  that the conditions (\ref{112})
and (\ref{113}) lead in second order to a Peierls-Boltzmann (P.B)
\cite{pei} form
\begin{multline}
\nu_{\bf{k}}\it{\phi_{\bf{k}} -
  \int\frac{M_{\bf{{\bf kjl}}}M_{\bf{jk,-l}}\phi_{\bf{k}}\phi_{\bf{j}}}{\omega_{\bf{k}}+\omega_{\bf{j}}+\omega_{\bf{l}}}{d^3\bf{j}} - \int\frac{M_{\bf{{\bf kjl}}}M_{\bf{lk,-j}}\phi_{\bf{k}}\phi_{\bf{l}}}{\omega_{\bf{k}}+\omega_{\bf{j}}+\omega_{\bf{l}}}{d^3\bf{j}}} -\\
\it{-
  \int\frac{|M_{\bf{{\bf kjl}}}|^2\phi_{\bf{j}}\phi_{\bf{l}}}{\omega_{\bf{k}}+\omega_{\bf{j}}+\omega_{\bf{l}}}{d^3\bf{j}} = D_{0{\bf k}}}
\label{114}
\end{multline}
where it turns out that the coefficients in the terms
$\phi_{\bf{k}}\it{\phi_{\bf{j}}}$ and
$\phi_{\bf{j}}\it{\phi_{\bf{l}}}$ have the effect after
integration of opposite signs, and those of
$\phi_{\bf{k}}\it{\phi_{\bf{j}}}$ and
$\phi_{\bf{k}}\it{\phi_{\bf{l}}}$ are equal. A similar equation
has been derived by Bouchaud and Cates \cite{bouch}. Note that the
value of $\left<h_{\bf k}h_{\bf j}h_{\bf l}h_{\bf m}\right>$ is given by
$\left<h_{\bf k}h_{\bf j}\right>\left<
h_{\bf l}h_{\bf m}\right>\;+\;\left<h_{\bf k}h_{\bf l}\right>\left<
h_{\bf m}h_{\bf j}\right>\;+\;\left<h_{\bf k}h_{\bf m}\right>\left< h_{\bf j}h_{\bf l}\right>$
\underline{and} several terms like $M^2\phi^3/\sum\omega$ plus
terms like $M^4\phi^4/\left(\sum\omega\right)^3$ so that in the
present theory, the four point correlation is not the sum of two
times two point correlations. One should note that within this
paper we are deriving an expansion rather than a closure. (A
systematic graphology for the higher terms is given in appendix A.
Feynman diagrams do not give the Peierls- Boltzmann equations
derived here.) Hence a Peierls-Boltzmann \cite {pei} structure has
emerged:
\begin{equation}
\nu_{\bf k}\phi_{\bf k}+\int\Lambda_{{\bf kj}}^{\left(1\right)}\phi_{\bf k}\phi_{\bf j}-\int\Lambda_{\bf{kj}}^{\left(2\right)}\phi_{\bf{k}-\bf{j}}\phi_{\bf j}=D_{0\bf{k}}
\label{115}
\end{equation}
A remarkably simple scaling argument emerges if we continue the
series (\ref{113}) and derive from it a series for the correlation
function $\phi_{\bf{k}}~=~\left< h_{\bf{k}}\it{h_{-{\bf
k}}}\right>$. The series amounts to a systematic expansion in
"(model - reality)". Each order in the expansion will have a
leading power in $\bf k$ which is $2a-\Gamma-2\mu+d$ greater than
the previous term where dimensionally
$M=k^a,\;\phi=k^{-\Gamma},\;\omega=k^{\mu},\;$ in $d$ dimensions,
because symbolically, the power series is in terms of
$M^2\phi/\omega^2$. In order that all terms have the same leading
power, it follows that
\begin{equation}
2a-\Gamma -2\mu +d=0 \;\mbox{and}\; \mu = \frac{2a+d-\Gamma}{2}
\label{116}
\end{equation}
A well known example arises in the Kolmogoroff dimensional
analysis of turbulence, caused by a source near $k=0$ (call this
case $KNS$). There, $d=3$, $a=1$ and Kolmogoroff argues that
$\Gamma =11/3$, so that the scaling argument gives $\mu =2/3$ (of
course Kolmogoroff invokes the dimensional argument to obtain
$2/3$, but our point is that there is a much more general argument
relating the time scale $\omega_{\bf{k}}^{-1}$ with the
correlation function $\phi_{\bf{k}}$). Note that the ease with
which the scaling argument can be checked to all orders in our
expansion confirms the value of the method. We can see that higher
terms in the expansion cannot alter powers in (\ref{116}), only
front factors.
 We develop an equation for $\omega$ below in (3.22), but it is
to be realised that whatever equation is deduced, all it gives is
a front factor; the power is determined by scaling. What is needed
is a transport equation that can naturally produce behaviour that
is more general than a decaying exponential, and to do this one
must treat time, or its Fourier transform $\omega$, as a natural
extension to four dimensions, i.e.
$\left<h_{\bf{k}\omega}h_{\bf{k}'\omega '}\right>$. To do this we
study the whole history distribution
$\mathcal{P}\left(\left[h_{\bf{k}\omega}\right]\right)$. Such
functions are of course well known in quantum mechanics after
their original introduction by Dirac \cite{dir}, in the form:
\begin{equation}
\exp\left(-\frac{i}{\hbar}\int\mathcal{L}d^3r dt\right)\, .
\label{117}
\end{equation}
Our self consistent method works for the extensions of
$\phi_{\bf{k}}$, $\omega_{\bf{k}}$ i.e. $\Phi_{\bf{k},\omega}$,
$\Omega_{\bf{k},\omega}$, and is presented in the next section.
Several papers are present on this problem in the literature
\cite{Ed64}-\cite{Frey96} but our method has the advantage of
producing simple equations (simple considering the complexity of
the problem) which allow us to produce explicit solutions due to
the ability to check scaling relations to all orders in a
systematic expansion.

\section{The Lagrangian formulation: a model}
\setcounter{equation}{0}
In this section, we study the simple case of a single degree of freedom obeying a noise driven linear equation
\begin{equation}
\frac{\partial h}{\partial t}+\nu h = f
\label{21}
\end{equation}
or in Fourier variables,
\begin{equation}
\left(i\omega+\nu\right)h_{\bf\omega}=f_{\bf\omega}.
\label{22}
\end{equation}

The main reason for considering such a simple model is that we aim at obtaining a first order differential equation in time for the non-linear systems we consider, that will match the static equations derived in our previous papers. This task is somewhat complicated by the fact that the noise in equation (\ref{21}) or in any physical system cannot be instantaneous, since it originates in physical processes. Consequently, in any physical system, the time derivative of, say, $\left<h\left(0\right)h\left(t\right)\right>$ at time $t=0$ is zero, and therefore a first order differential evolution equation cannot evolve the system in time. To understand what is going on and to obtain the correct matching condition, we study the system (\ref{21}-\ref{22}) by considering a non instantaneous noise described by the correlation:
\begin{equation}
\left<f_{\omega}f_{-\omega}\right>=\frac{H}{\pi}\left[\omega^2+l^2\right]^{-1}
\label{23}
\end{equation}
so that
\begin{equation}
\left<h_{\omega}h_{-\omega}\right>=\frac{H/\pi}{\left[\omega^2+\nu^2\right]\left[\omega^2+l^2\right]},
\label{24}
\end{equation}
\textit{i.e.}
\begin{equation}
\Phi\left(t\right)=\left<h\left(0\right)h\left(t\right)\right>=\frac{H}{l^2-\nu^2}\left(\frac{{\rm e}^{-\nu|t|}}{\nu} - \frac{{\rm e}^{-l|t|}}{l}\right).
\label{25}
\end{equation}
Then
\begin{equation}
\Phi\left(0\right)=\frac{H}{\left(l+\nu\right)l\nu}
\label{26}
\end{equation}
and, as expected,
\begin{equation}
\dot{\phi}\left(0\right)=0.
\label{27}
\end{equation}
However, if $l>>\nu$, \textit{i.e.} almost instantaneous noise,
\begin{equation}
\Phi\left(0\right)=\frac{H}{l^2\nu}\equiv\frac{D_0}{\nu}
\label{28}
\end{equation}
and
\begin{equation}
\dot{\Phi}\left(\tau\right)=-D_0,
\label{29}
\end{equation}
for $1/l\;<<\;\tau\;<< 1/\nu.$ This is described in figure 1

\textbf{SHOW FIGURE 1 HERE}.

In the limit $l\rightarrow\infty$, the Fokker-Planck equation for the probability of finding $h$ at $t$ satisfies:
\begin{equation}
\left[\frac{\partial}{\partial t}-\frac{\partial}{\partial h}\left(D_0\frac{\partial}{\partial h}+\nu h\right)\right]P=0,
\label{210}
\end{equation}
which gives
\begin{equation}
\frac{\partial \Phi}{\partial t}+\nu\Phi=0 \;\;\;\; t>0
\label{211}
\end{equation}
and
\begin{equation}
-\frac{\partial \Phi}{\partial t}+\nu\Phi=0 \;\;\;\; t<0
\label{212}
\end{equation}
and $\Phi\left(0\right)=D_0/\nu$. The awkwardness of (\ref{28})-(\ref{212}) is removed by putting in the full dependence on $l$, but more simply, as described above, confining ourselves to $t>0$,we have the first order differential equation (\ref{211}) with the initial condition $\Phi\left(0\right)=D_0/\nu$, that implies a finite derivative $\dot{\Phi}\left(0\right)=-D_0$ at $t=0$. This matches the static equation obtained from the Fokker-Planck equation (\ref{210})
\begin{equation}
D_0-\nu\left<hh\right>=0
\label{213}
\end{equation}
The form of the Fourier transform of $\Phi\left(t\right)$, $\Phi\left(\omega\right)$, suggests the structure of $\Phi_{k,\omega}$ in the non-linear field theory.

For the simple linear case,
\begin{equation}
\Phi_{\omega}=\frac{H/\pi}{[\omega^2+\nu^2][\omega^2+l^2]}\equiv\frac{H/\pi}{\Omega^*\Omega},
\label{214}
\end{equation}
where $\Omega =[i\omega +\nu][i\omega +l]$.

Thus the two decays of $\Phi (t)$, for $t>0$, are present as zeroes of $\Omega$ (poles in $\Phi_{\omega}$, in the upper half of the complex $\omega$ plane.) In the limit $l\rightarrow\infty$, the situation is similar, but there is only a single decay
\begin{equation}
\Phi_{\omega}=\frac{D_0}{\omega^2+\nu^2}=\frac{D_0/\pi}{\Omega^*\Omega},
\label{215}
\end{equation}
where $\Omega =[i\omega +\nu]$.

The natural model to try for the $\Phi$ of the non-linear equation is \cite{Ed64}
\begin{equation}
\Phi_{{\bf k}\omega}=\frac{D_{{\bf k}\omega}}{\Omega_{{\bf k}\omega}\Omega^{*}_{{\bf k}\omega}},
\label{216}
\end{equation}
where $D_{{\bf k}\omega}=D_{-{\bf k},-\omega}>0$ with no singularities in the complex $\omega$ plane,
and where $\Omega$ gives in the first self consistent approximation a simple decay described by
\begin{equation}
\Omega_{{\bf k}\omega}=i\omega + \omega_{\bf{k}} \, .
\label{217}
\end{equation}
However, there will be a much more complicated time dependence in the full $\Phi$ than one simple decay.
(Equations (\ref{214}) and (\ref{215}) form a simple example where the decay is given to first approximation
by a simple decay but indeed the behaviour is more complicated)

One possible definition of $\Omega$ is to use the response function and define $\Omega$ by
\begin{equation}
\frac{D_0}{\Omega_{{\bf k}\omega}}=\left<f_{-{\bf k}-\omega}h_{{\bf k}\omega}\right>
\label{218}
\end{equation}
and employed in mode-mode coupling studies \cite{bouch,kaw,moore}.
We will take the view that we can write $\Phi$ in terms of a sum
of exponential decays which can be extended to cover continuous
distributions \textit{i.e.} branch cuts rather than simple poles.
Confining ourselves to poles for the moment, if we write
\begin{equation}
\Phi_{{\bf k}\omega}=\sum_{\bf l}\frac{A_{l,{\bf k}}}{\omega^2+\omega^{2}_{{\bf k},{\bf l}}}=\frac{D_{{\bf k}\omega}}{\Pi_{\bf l}\left(\omega^2+\omega^{2}_{{\bf k},{\bf l}}\right)},
\label{219}
\end{equation}
we find that $D_{{\bf k}\omega}$ is an even function in $\omega$ (a polynomial for a finite sum of simple decays), and it has no singularities.

Thus
\begin{equation}
\Omega_{{\bf k}\omega}=\frac{1}{\prod_{\bf l}\left(i\omega+\omega_{{\bf k},{\bf l}}\right)}
\label{220}
\end{equation}
and
\begin{equation}
\Omega^{*}_{{\bf k}\omega}=\frac{1}{\prod_{\bf l}\left(-i\omega+\omega_{{\bf k},{\bf l}}\right)}
\label{221}
\end{equation}

When $\Phi_{{\bf k}}\left(t\right)$ is obtained from (\ref{219}) only the poles $\omega=i\omega_{{\bf k},l}$ contribute to $t>0$ and their conjugates for $t<0$.

The strategy we will adopt is to construct a transport equation for $\Phi_{{\bf k}\omega}$ in which the first order approximation (\ref{217}) for $\Omega_{{\bf k}\omega}$ will be useful and it will result in a higher order approximation for $\Phi_{{\bf k}\omega}$ and consequently for $\Phi_{{\bf k}}\left(t\right)$ that will show now a decay that is much more complicated than a simple exponential.

\section{The expansion}
\setcounter{equation}{0}
The starting point is equation (\ref{15}), where in order to make our notation less cumbersome,
we denote the $d+1$ vector $({\bf k},\omega )$ by ${\bf K}$ and write the equation in the form
\begin{equation}
\Omega_{0{\bf K}}h_{{\bf K}}+\sum M_{{\bf {\bf kjl}}}h_{{\bf J}}h_{{\bf L}}-f_{{\bf K}}={\bf 0}
\label{31}
\end{equation}
We define next $P\left\{h_{{\bf K}};f_{{\bf K}}\right\}$ to be the distribution of the $h_{{\bf K}}$'s
in the presence of a given noise, $f_{{\bf K}}$. $P$ is to be averaged eventually over the noise.
The Liouville equation (\ref{16}) is now replaced by
\begin{equation}
\left[\Omega_{0{\bf K}}h_{{\bf K}}+\sum_{{\bf JL}}M_{{\bf {\bf kjl}}}h_{{\bf J}}h_{{\bf L}}-f_{{\bf K}}\right]P=0
\label{32}
\end{equation}
which is similar to a Fermi supplementary condition. Equivalently,
equation (\ref{32}) can be replaced by:
\begin{equation}
\sum_{{\bf K}}\frac{\partial}{\partial h_{{\bf K}}}\left(\Omega_{0{\bf K}}h_{{\bf K}}+
\sum_{{\bf JL}}M_{{\bf {\bf kjl}}}h_{{\bf J}}h_{{\bf L}}-f_{{\bf K}}\right)P=0,
\label{33}
\end{equation}
to obtain the correct hierarchy of field correlations. This is achieved by multiplying the above
equation by products like $h_{{\bf L}1}...h_{{\bf L}N}$ and integrating by parts.
A simple example is obtained by multiplication by $h_{{\bf K}}$ that yields the correct average of equation (\ref{31}).
\begin{equation}
\Omega_{0{\bf K}}\left<h_{{\bf K}}\right>+\sum_{{\bf J},{\bf L}}M_{{\bf {\bf kjl}}}\left<h_{{\bf J}}h_{{\bf L}}\right>-f_{{\bf K}}=0.
\label{34}
\end{equation}
An alternative derivation of equation (\ref{33}) starts with consideration of a $d+2$ dimensional system in
which the equation of the form
\begin{equation}
\frac{\partial}{\partial s}h_{{\bf K}}(s)=\left[\Omega_{0{\bf K}}h_{{\bf K}}(s)+
\sum_{{\bf JL}}M_{{\bf JKL}}h_{{\bf J}}(s)h_{{\bf L}}(s)-f_{{\bf K}}+g_{{\bf K}}(s)\right]
\label{35}
\end{equation}
is considered, where $g_{{\bf K}}(s)$ is a $d+2$ dimensional noise obeying
\begin{equation}
<g_{{\bf K}}(s)>=0 \; and \; <g_{{\bf K}}(s)g_{-{\bf K}}(s')>={\mathcal D}_{{\bf K}}\delta (s-s').
\label{36}
\end{equation}
(Note that $f_{{\bf K}}$ does not depend on $s$. Therefore it plays the role of quenched randomness).
The Fokker-Planck equation for ${\mathcal P}\left\{h_{{\bf K}},s\right\}$, the distribution of a given
configuration $\left\{h_{{\bf K}}\right\}$ is given by
\begin{equation}
\frac{\partial {\mathcal P}}{\partial s}=\sum_{{\bf K}}\frac{\partial}{\partial h_{{\bf K}}}\left[{\mathcal D}_{{\bf K}}
\frac{\partial}{\partial h_{-{\bf K}}}+\Omega_{0{\bf K}}h_{{\bf K}}+\sum_{{\bf JL}}M_{{\bf {\bf kjl}}}h_{{\bf J}}
h_{{\bf L}}-f_{{\bf K}}\right]{\mathcal P}.
\label{37}
\end{equation}
The 'steady state' ($s$ independent) $\mathcal P$ in the limit where ${\mathcal D}_{{\bf K}}$ is zero is the
distribution of $h_{{\bf K}},P$ and equation (\ref{33}) is recovered.

In order to construct an expansion for $P$, we write equation (\ref{33}) as:
\begin{multline}
\sum_{{\bf K}}\frac{\partial}{\partial h_{\bf K}}\left(\frac{{\mathcal D}_{{\bf K}}}{\Omega_{-{\bf K}}}
\frac{\partial}{\partial h_{\bf -K}}+\Omega_{{\bf K}}h_{{\bf K}}\right)+\sum_{{\bf K}}\frac{\partial}{\partial
h_{{\bf K}}}\left[\sum_{{\bf JL}}M_{{\bf {\bf kjl}}}h_{{\bf J}}h_{{\bf L}}-f_{{\bf K}}\right]\\
+\sum_{{\bf K}}\frac{\partial}{\partial h_{{\bf K}}}\left[\left(\Omega_{0{\bf K}}-\Omega_{{\bf K}}\right)h_{{\bf K}}-
\frac{{\mathcal D}_{{\bf K}}}{\Omega_{-{\bf K}}}\frac{\partial}{\partial h_{-{\bf K}}}\right]P=0
\label{38}
\end{multline}
Notice that since the sum is over all $\bf K$, both $\Omega_{{\bf K}}$ and $\Omega_{-{\bf K}}$ will appear in the
second derivative.

We expect an expansion, of the average of $P$ over the noise, around the Gaussian
\begin{equation}
P_0\propto{\rm exp}\left[-\frac{1}{2}\sum_{{\bf K}}\frac{h_{{\bf K}}h_{-{\bf K}}}{\Phi_{{\bf K}}}\right]
\label{39}
\end{equation}
and, as before, associate in (\ref{38}) a notional $\lambda$ to
the second term on the left hand side of equation (\ref{39}) and
$\lambda^2$ to the third term. Expanding to second order in the
$\lambda$ "Chapman and Enskog" expansion, we get
\begin{equation}
P=P_0\left(1-\sum_{{\bf {\bf kjl}}}\frac{M_{{\bf {\bf kjl}}}h_{{\bf J}}h_{{\bf L}}h_{-{\bf K}}}{\left[\Omega_{{\bf J}}+
\Omega_{{\bf L}}+\Omega_{-{\bf K}}\right]\Phi_{{\bf K}}}+...\right)
\label{310}
\end{equation}

The condition that $<h_{{\bf K}}h_{-{\bf K}}>$ calculated to second order in $\lambda$ is equal to $\Phi_{{\bf K}}$
(that is the zero order result) yields an equation for $\Phi$ in terms of $\Omega$.
\begin{multline}
\frac{1}{2}\left(\Omega_{0{\bf K}}+\Omega_{0,{\bf -K}}\right)\Phi_{{\bf K}}-2\sum_{{\bf JL}}\frac{M_{{\bf {\bf kjl}}
}M_{{\bf JK,-L}}}{\left[\Omega_{{\bf J}}+\Omega_{{\bf L}}+\Omega_{-{\bf K}}\right]}\Phi_{{\bf L}}\Phi_{{\bf K}}-\\
\sum_{{\bf JL}}\frac{|M_{{\bf {\bf kjl}}}|^2}{\left[\Omega_{{\bf J}}+\Omega_{{\bf L}}+\Omega_{-{\bf K}}\right]}
\Phi_{{\bf J}}\Phi_{{\bf L}}=\frac{D_0}{2}\left[\frac{1}{\Omega_{{\bf K}}}+\frac{1}{\Omega_{-{\bf K}}}\right].
\label{311}
\end{multline}
We can recover the structure of a transport equation, familiar
from the static case, \textit{i.e.} "un-pick" $\Omega_{0{\bf
K}}+\Omega_{0,-{\bf K}}$ and $1/\Omega_{{\bf K}} + 1/\Omega_{-{\bf
K}}$ by returning to equation (\ref{31}), multiplying it by
$h_{-{\bf K}}$ and averaging over the distribution (\ref{310}).
Using also equation (\ref{218}) we obtain
\begin{equation}
\Omega_{0{\bf K}}\Phi_{{\bf K}}-2\sum_{{\bf JL}}\frac{M_{{\bf {\bf kjl}}}M_{{\bf JK,-L}}}{\left[\Omega_{{\bf J}}+
\Omega_{{\bf L}}+\Omega_{-{\bf K}}\right]}\Phi_{{\bf L}}\Phi_{{\bf K}}-\sum_{{\bf JL}}
\frac{|M_{{\bf {\bf kjl}}}|^2}{\left[\Omega_{{\bf J}}+\Omega_{{\bf L}}+\Omega_{-{\bf K}}\right]}
\Phi_{{\bf J}}\Phi_{{\bf L}}=\frac{D_0}{\Omega_{-{\bf K}}}.
\label{312}
\end{equation}
At this point we use, just in the non linear term, the first order approximation
$\Omega_{{\bf K}}=i\omega+\omega_{{\bf K}}$. This has the advantage that now
$\Omega_{{\bf J}}+\Omega_{{\bf L}}+\Omega_{-{\bf K}}=\omega_{\bf j}+\omega_{\bf l}+\omega_{\bf k}$.
Fourier transforming back and recalling that the zeroes of $\Omega_{-{\bf K}}$ are in the lower complex
$\omega$ plane, we find for $t>0$ the local equation
\begin{multline}
\frac{\partial\Phi_{\bf k}(t)}{\partial t}+\nu_{\bf k}\Phi_{\bf k}(t)-2\sum\frac{M_{{\bf {\bf kjl}}}M_{\bf{jk},-\bf{l}}}{\omega_{\bf j}+\omega_{\bf l}+
\omega_{\bf k}}\phi_{\bf l}\Phi_{\bf k}(t)-\\
\sum\frac{|M_{\bf{{\bf kjl}}}|^2}{\omega_{\bf j}+\omega_{\bf l}+\omega_{\bf k}}\Phi_{\bf j}(t)\Phi_{\bf l}(t)=0
\label{313}
\end{multline}
and
\begin{multline}
-\frac{\partial\Phi_{\bf k}(t)}{\partial t}+\nu_{\bf k}\Phi_{\bf k}(t)-2\sum\frac{M_{{\bf kjl}}M_{\bf{jk},-\bf{l}}}{\omega_{\bf j}+\omega_{\bf l}+
\omega_{\bf k}}\phi_{\bf l}\Phi_{\bf k}(t)-\\
\sum\frac{|M_{{\bf kjl}}|^2}{\omega_{\bf j}+\omega_{\bf l}+\omega_{\bf k}}\Phi_{\bf j}(t)\Phi_{\bf l}(t)=0\;\mbox{for}\;t<0.
\label{314}
\end{multline}
Notice that the above is possible because the coefficients, $M_{{\bf {\bf kjl}}}$, do not depend on the fourth
components of the vector. The initial conditions with which equation (\ref{313}) has to be solved are
\begin{equation}
\Phi_{\bf k}(0)=\phi_{\bf k},\quad\mbox{for all k,}
\label{315}
\end{equation}
where $\phi_{\bf k}$ is the static correlations. The static equation determining $\phi_{\bf k}$ is
\begin{equation}
\nu_{\bf k}\phi_{\bf k}-2\sum\frac{M_{{\bf kjl}}M_{\bf{jk},-\bf{l}}}{\omega_{\bf j}+\omega_{\bf l}+\omega_{\bf k}}\phi_{\bf l}\phi_{\bf k}-\sum\frac{|M_{{\bf kjl}}|^2}{\omega_{\bf j}+
\omega_{\bf l}+\omega_{\bf k}}\phi_{\bf j}\phi_{\bf l}=_{{\bf k}0}.
\label{316}
\end{equation}
It can be shown that
\begin{equation}
\lim_{t\to 0+}\frac{\partial \Phi_{\bf k}(t)}{\partial t}=-_{{\bf k}0},
\label{317}
\end{equation}
is an exact relation, obeyed by the exact two point function.
Using this general result, we see that as $t\rightarrow 0+$, the
evolution equation for $\Phi_{\bf k}(t)$, eq. (\ref{313}) fits exactly
onto the static equation for $\phi_{\bf k}$, eq. (\ref{316}). The
equations (\ref{313}) and (\ref{316}) were originally derived for
the driven Navier-Stokes equations \cite{Ed64}, but although
equation (\ref{313}) can give the Kolmogoroff \cite{McComb}
spectrum with a good value for the front factor, the boundary
condition was not understood at that time, and this hindered
further development of this approach to non-linear equations by
this route for several decades. Note that the simplicity of the
basic equations (\ref{313}-\ref{315}) and (\ref{316}) sets this
approach apart from mode-mode coupling theories. We have found a
plausible way (the structure of $\Omega$) which leads to these
manageable and transparent equations. The amazing feature of the
self-consistent approach is that the time-dependent equation has
an explicit and local dependence in time. It now offers a way to
complete the system of functions, $\phi_{\bf{k}}$,
$\omega_{\bf{k}}$ in a satisfying way. To the best of our
knowledge the direct evaluation of the indices of $\phi_{\bf{k}}$
and  $\omega_{\bf{k}}$ and the universal \underline{structure} of
the time dependent correlation functions are not available from
other treatments. The simple structure of our equations  now
offers a way to complete the system of functions,
$\phi_{\bf k},\omega_{\bf k}$, in a satisfying way. We define $\omega_{\bf k}$
customarily to be given by
\begin{equation}
\omega_{\bf k}^{-1}=\frac{\int_0^\infty\Phi_{\bf k}(t)\;{\rm d}t}{\phi_{\bf k}},
\label{318}
\end{equation}
which is a natural definition, if we think about a single mode
decay.
  Integrating equation (\ref{313}) over time, taking into account (\ref{317})
and the fact that $\int_0^{\infty}\frac{\partial
  \Phi_{\bf k}(t)}{\partial t}\;{\rm d}t=-\phi_{\bf k}$, we obtain
\begin{multline}
\omega_{\bf k}=\nu_{\bf k}-2\sum_{j,l}\frac{M_{{\bf kjl}}M_{\bf{jk},-\bf{l}}}{[\omega_{\bf j}+\omega_{\bf l}+\omega_{\bf k}]}\phi_{\bf l}-\sum\frac{|M_{{\bf kjl}}|^2}{[\omega_{\bf j}+\omega_{\bf l}+\omega_{\bf k}]}\times\\
\frac{\phi_{\bf j}\phi_{\bf l}}{\phi_{\bf k}}\left(\frac{\omega_{\bf k}}{\omega_{\bf{j},\bf{l}}}\right)\,
, \label{319}
\end{multline}
where
\begin{equation}
\int_0^{\infty}\Phi_{\bf l}(t)\Phi_{\bf j}(t)\;{\rm
d}t=\frac{\phi_{\bf l}\phi_{\bf j}}{\omega_{\bf{j},\bf{l}}} \, .
\label{320}
\end{equation}
Neglecting $D_{0k}$ and $\nu_{\bf k}$ in equation (\ref{316}) (the
static equation) and (\ref{319}) (the $\omega$ equation) we find
that in the inertial range
\begin{equation}
\omega_{\bf k}\,=\,\sum\,\frac{|M_{{\bf kjl}}|^{2}\,\phi_{\bf j}\phi_{\bf l}\,(\omega_{\bf{j},\bf{l}}\,-\,\omega_{\bf k})}{(\omega_{\bf k}\,+\,\omega_{\bf j}\,+\,\omega_{\bf l})\,\phi_{\bf k}\,\omega_{\bf{j},\bf{l}}}
\label{321}
\end{equation}
A similar equation has been derived by a different method by
Edwards and McComb \cite{Ed69}, who used it to derive the
Kolmogoroff front factor, achieving a reasonable value. Details of
the alternative method are given in the book of McComb
\cite{McComb}. It is easy to see that  eq.(\ref{319}) leads just
to the scaling relation discussed in the introduction because
$\omega_{\bf{j},\bf{l}}$ scales as $\omega$. This scaling relation together
with the static equation gives

\parbox{11cm}{
\begin{eqnarray*}
\omega_{\bf k}&=&B_1k^{3/2}\;\mbox{in KPZ in 1+1 D}\\
&=&B_2k^{1.7}\;\mbox{in KPZ in 2+1 D [2]}\\
\omega_{\bf k}&=&B_3k^{2/3}\;\mbox{in KNS [13]}
\end{eqnarray*}}\hfill
\parbox{1cm}{\begin{eqnarray}\end{eqnarray}\label{322}}

\section{A closer look at the steady state in the inertial range}
\setcounter{equation}{0}
The structure of the steady state equation has the form
\begin{equation}
\int\Lambda^{(1)}(j,k)\phi_{\bf j}\phi_{\bf k}\;{\rm d}^dj-\int\Lambda^{(2)}(j,k)\phi_{\bf j}\phi_{\bf{k}-\bf{j}}\;{\rm d}^dj=_{{\bf k}0}-\nu_{\bf k}\phi_{\bf k}.
\label{41}
\end{equation}
The kernel $\Lambda^{(2)}$ stemming from the $|M_{{\bf kjl}}|^2$ term is
positive definite but $\Lambda^{(1)}(j,k)$ may attain positive as well
as negative values depending on the specific problem and the values of
$j$ and $k$. For $KPZ$ with noise, that is white (in space) it was
shown explicitly [10] that the exponent characterising the leading
small $k$ dependence is obtained by equating the left hand side of
eq. (\ref{41}) to zero. The concept of inertial range is given thus a
definite meaning since the exponent does not depend neither on the
source nor the 'viscosity'. For $KPZ$ with noise given by
$_{{\bf k}0}=D_0k^{-2\rho}$ \cite{katz} it was shown that up to some
threshold $\rho_0$ the inertial range still exists and the exponent
 characterising the small $k$ behaviour does not depend on the source
 or the 'viscosity' (and therefore, its value is the same as in the white noise case).
 For $\rho>\rho_0$, the system is driven by the noise and even the leading $k$ dependence
 depends on the source term. The problem of turbulence seems to combine both behaviours in an
interesting way. The simplest conceptual picture of the inertial
range is that offered by Kolmogoroff, where random forces put
energy into a fluid at low $\bf{k}$, and viscosity removes it at
high $\bf{k}$. We can make an extreme model of this by taking the
input to be $\varepsilon\delta\left({\bf{k}}\right)$ and the
output to be
$\varepsilon\frac{\delta\left(\left|\bf{k}\right|-\infty\right)}{4\pi\left|\bf{k}\right|^2}$
which we write symbolically as
$\varepsilon\delta\left(k-\infty\right)$, so that
\begin{equation}
\int\Lambda^{\left(1\right)}\phi_{{\bf{k}}}\phi_{{\bf{j}}}\;d^3{\bf{j}}-\int\Lambda^{\left(2\right)}\phi_{{\bf{k-j}}}\phi_{{\bf{j}}}\;d^3{\bf{j}}=\varepsilon\left[\delta\left(k\right)-\delta\left(k-\infty\right)\right].
\label{42}
\end{equation}
The Navier Stokes $\mathcal{M}$ guarantees that the integral over $\bf{k}$ of the left hand side vanishes for any $\phi_{\bf k}$
\begin{equation}
\int\left(\int\Lambda^{\left(1\right)}\phi_{{\bf{k}}}\phi_{{\bf{j}}}\;d^3{\bf{j}}-\int\Lambda^{\left(2\right)}\phi_{{\bf{k-j}}}\phi_{{\bf{jh}}}\;d^3{\bf{j}}\right)\;d^3{\bf{k}}=0
\label{43}
\end{equation}
and so matches
\begin{equation}
\int
d^3{\bf{k}}\;\varepsilon\left[\delta\left(k\right)-\delta\left(k-\infty\right)\right]=0
\, . \label{44}
\end{equation}
To clarify (\ref{42})-(\ref{44}) we offer a model in Appendix B.

We have computed the value of the left hand side of equation (\ref{42})
for a range of values of $\Gamma$, and calling the quantity
$Z\left(\Gamma\right)$, we find that the value of $\Gamma$ for which
$Z\left(\Gamma\right)=0$ is (see figure 2)
\begin{equation}
\Gamma=3.6667
\label{415}
\end{equation}
to the accuracy of our computation (For $KPZ$, we have already found the value 2.59).

\begin{center}
show figure 2 here
\end{center}

We have used the $\phi_{\bf k}=k^{-11/3}$ and integrated the left hand side of eq.(\ref{42}) over a finite $k$ sphere, interchanging the $k$ and $j$ integration, we obtain, as expected, a non-zero result, in spite of the fact that for any finite $k$ the integrand is zero.

This results gives one the confidence to proceed to the much more difficult problem of the time dependence.

It is interesting to consider briefly another example which is Navier-Stokes driven white noise (call it $WNS$). In the problem, the viscosity can again only influence very large $k$, but now the source plays a vital part. Integrals converge, with the solution of
\begin{equation}
\int\Lambda^{(1)}\phi_{\bf k}\phi_{\bf j}-\int\Lambda^{(2)}\phi_{\bf{k}-\bf{j}}\phi_{\bf j}=D_0
\label{416}
\end{equation}
where $D_0$ is now a constant, and one finds

\parbox{11cm}{
\begin{eqnarray*}
\phi_{\bf k}&\propto&Ek^{-5/3}D_0^{2/3},\; \mbox{and}\\
\omega_{\bf k}&\propto&Fk^{5/3}D_0^{1/3}\;\\
\end{eqnarray*}}\hfill
\parbox{1cm}{\begin{eqnarray}\end{eqnarray}\label{417}}

If one tried to solve this in the $KPZ$ style of ignoring $D_0$, one would of course get Kolmogoroff again, but Kolmogoroff's $-11/3$, $2/3$ regime will not satisfy equation(\ref{416}).

\section{The time dependent correlation function}
\setcounter{equation}{0}
So far we have obtained the solution for the steady state correlation function,
$\left<h_{{\bf{k}}}\left(t\right)h_{{\bf{-k}}}\left(t\right)\right>$ in the two cases of $KPZ$ and $NS$.
 Turning to the time dependent function, $\Phi_{{\bf{k}}}\left(t\right)$, at first sight we are faced with a
 formidable set of equations, given by (\ref{313}), (\ref{316}) and (\ref{319}), of which (\ref{313}) is
 central.We give firstly a crude argument and relegate a full discussion to Appendix C. If we write
 $\Phi_{\bf k}~=~\phi_{\bf k}e^{-\Gamma_{\bf k}(t)}$ we can manipulate
 (\ref{313}) and (\ref{41}) to obtain
\begin{equation}
\frac{\partial \Gamma_{\bf k}}{\partial t}\,=\,\int \, \Lambda_{2}\,
\frac{\phi_{\bf j}\,\phi_{\bf{k}-\bf{j}}}{\phi_{\bf k}}\,\Big(1\,-\,e^{-\Gamma_{\bf j}(t)-\Gamma_{\bf{k}-\bf{j}}(t)+\Gamma_{\bf k}(t)}\Big)\,\mbox{d}^{3}j
\, ,
\end{equation}
and argue (to be justified at greater length in Appendix C) the
solution is dominated by the that part of $j$ space where

\begin{equation}
1\,=\,e^{-\Gamma_{\bf j}-\Gamma_{\bf{k}-\bf{j}}+\Gamma_{\bf k}}
\end{equation}
i.e. $\Gamma_{\bf j}(t)+\Gamma_{\bf{k}-\bf{j}}(t)~=~\Gamma_{\bf k}(t)$.
We can scale out $t$ and have
\begin{equation}
\Gamma_{\bf j}+\Gamma_{\bf{k}-\bf{j}}\,=\,\Gamma_{\bf k} \label{53a}
\end{equation}
where $\Gamma_{\bf k}~=~\Gamma_{\bf k}(1)$ etc.. Consider this guess:
$\Gamma_{\bf k}~\sim ~k^{2}$. Then we would need
$k^{2}~=~(\bf{k}+\bf{j})^{2}+j^{2}$ which by Pythagoras's theorem
gives the locus of $j$ to be a sphere of diameter $k$. Inside the
sphere
\begin{equation}
\Gamma_{\bf k}\,>\,\Gamma_{\bf{k}-\bf{j}}\,+\,\Gamma_{\bf j}
\end{equation}
Outside the sphere
\begin{equation}
\Gamma_{\bf k}\,<\,\Gamma_{\bf{k}-\bf{j}}\,+\,\Gamma_{\bf j}
\end{equation}
so the guess is inadequate. The only way to find a region of $j$
space wherein (\ref{53a}) obtains is when $\Gamma$ is linear when
$\bf{j}$ equals $\alpha \bf{k}$ such that
$1~\geqslant~\alpha~\geqslant~0$. The integral goes outside that
locus, but the corrections are
logarithmic and derived in Appendix C. Returning to
$\Gamma_{\bf k}(t)$
\begin{equation}
\Gamma_{\bf k}(t)\,=\,a\,|\bf{k}|\,t^{1/\mu}
\end{equation}
where $a$ is a constant of the appropriate dimensions. For $\mu
~>~1$, we derive in Appendix C the long time decay of
$\Phi_{\bf k}(t)$, including the logarithmic corrections
\begin{equation}
\Phi_{\bf k}(t)\,\varpropto\,\Big(a\,|\bf{k}|\,t^{1/\mu}\Big)^{\frac{d-1}{2}}\,exp(-a\,|\bf{k}|\,t^{1/\mu})
\, .
\end{equation}
\newpage
\section{Some speculations}
\setcounter{equation}{0}
Working to a kernel of order $M^2$, the "transport" equation we have derived has the classic form with operators $A$, $B$ such that
\begin{equation}
A\phi + B\phi\phi = D;
\label{61}
\end{equation}
but the solutions for different physical cases seem radically different.

For $KPZ$, it appears that $A$ and $D$ are not central to the inertial range away from $k=0$ and $k=\infty$, and the solution is given by taking a power law $k^{-\Gamma}$, and deriving in detail
\begin{equation}
B\phi\phi=Z\left(\Gamma\right)
\label{62}
\end{equation}
so that the value of $\Gamma$ must be such that
\begin{equation}
Z\left(\Gamma\right)=0
\label{63}
\end{equation}
and this gives the solution in the inertial range. The solution is then improved by adding the effects of $A\phi$ and $D$. However, for problems rich in dimensional parameters, e.g. instead of $M_{{\bf{{\bf kjl}}}}$ being a combination of powers, it could be a function like $\rm{e}^{-\left(k^2+\left({\bf{k}}+{\bf{j}}\right)^2+j^2\right)c^2}$, containing a new constant $c$ with dimensions (length), likewise for $A$ and $D$, then equation \ref{61} is more of a standard Boltzmann form and, for example, a perturbation approach would be
\begin{equation}
\phi=\frac{D}{A}-B\frac{D}{A}\frac{D}{A}+\cdots
\label{64}
\end{equation}
which can be improved in various ways. Navier-Stokes seems to have aspects of both of these, for at first sight, if one tries a power, then $B\phi\phi=0$ has the Kolmogoroff solution derived just as the $KPZ$ is derived. In fact the lack of uniform convergence puts Kolmogoroff into the form:
\begin{equation}
B\phi\phi=D
\label{65}
\end{equation}
and the $A\phi$ term only comes in to polish the solution at high $k$. For $WNS$ an attempt to use (\ref{63}) inevitably gives the Kolmogoroff indices which are quite incorrect for this problem. One must use (\ref{65}) since $D$ is a constant for all $\bf k$, not just being non-zero at $k=0$ as in $KNS$.

Suppose now that we proceed to the next orders, which will give
\begin{equation}
A\phi+B\phi\phi+C\phi\phi\phi+E\phi\phi\phi\phi+\cdots =D
\label{66}
\end{equation}
Can we now try $\phi\sim k^{-\Gamma}$ and obtain
\begin{equation}
Z_B\left(\Gamma\right)+Z_C\left(\Gamma\right)+Z_E\left(\Gamma\right)+\cdots =0.
\label{67}
\end{equation}

For $KPZ$, $Z_B\left(\Gamma\right)=0$ gives a $\Gamma$ which
cannot be obtained from dimensional analysis. More detail is given
in Appendix D. There is no reason to believe that if we proceeded
to equation (\ref{67}) it will not give again a $\Gamma$, not
quite the same as that of equation (\ref{62}), but if the self
consistent approach works well it will be close. Certainly
numerical simulations equivalent to all orders in (\ref{66}) do
agree remarkably well with the second order, (\ref{62}). However,
in other equations, in particular $KNS$, the deficiency in
dimensions of the equations leads to the Kolmogoroff solution both
by looking at $Z_B\left(\Gamma\right)=0$, and by balancing the
source term with the energy cascade. What happens if we go to the
higher orders in $KNS$? Could it be that equation (\ref{67}) has a
non Kolmogoroff solution? The calculation of $C,E$ etc. in
equation (\ref{66}) is a formidable undertaking and, unless some
new dimensional quantity is needed to make $C,E$ convergent, there
seems to no reason that Kolmogoroff should not again be the
solution.
\section{Acknowledgements}
S.F.E. acknowledges with gratitude the award of a senior research
fellowship from the Leverhulme Foundation, and for support from
the Polymers and Colloids Group at the Cavendish Laboratory, and
from the Sackler Foundation at Tel Aviv University.

M.S. acknowledges support from the Sackler Foundation at the
Sackler Institute at Tel Aviv University, and the Polymers and
Colloids Group at the Cavendish Laboratory, Cambridge University.

\newpage
\section{Appendices}
\appendix
\section{}
\setcounter{equation}{0}
 In quantum field theory the algebraic series which, in for example the basic papers
of Schwinger \cite{schwinger}, can be rather impenetrable, becomes
much simpler with the use of Feynman diagrams. In this paper, we
are proposing that the Boltzmann equation is the correct target
for equations like $KPZ$, we should offer a graphology to simplify
the appreciation of the formal expansion. This has in fact been
done in the original study of $NS$ many years ago \cite{Ed64,
McComb}, and extended to many body problems by Sherrington
\cite{sher}, and then fully extended to turbulence problems
\cite{quian}. However, it is not well known and so we reproduce it
here, and extend the analysis to some new cases, in particular the
expressions for fourth moments. We use the notation of the steady
state problem, although it works equally for the time dependent
case. The problem is to find $P$ which satisfies (\ref{18}). $P$
is to be expanded about $P_0$, which satisfies (\ref{19}), and
this is effected by introducing $D$ and $\omega$, so that writing
\begin{eqnarray}
\Delta D &=&D-D_0\label{71}\\
\mbox{and}\;\;\Delta\omega&=&\omega-\nu\label{72}
\end{eqnarray}
\begin{equation}
\frac{\partial}{\partial h}\left(D\frac{\partial}{\partial h}+\omega h-\Delta D\frac{\partial}{\partial h}-\Delta\omega h + Mhh\right)\left(P_0+P_1+P_2\cdots\right)=0
\label{73}
\end{equation}
and ascribing $"\lambda"$ to $Mhh$ and $"\lambda^2"$ to $\Delta D$ and $\Delta\omega$, we expand

\begin{eqnarray}
P &=& P_0+P_1+P_2+\cdots\\ \nonumber
P_1 &=& G\frac{\partial}{\partial h}MhhP_0\\ \nonumber
P_2 &=& G\frac{\partial}{\partial h}MhhG\frac{\partial}{\partial h}MhhP_0\\ \nonumber
&+&G\bigg(\frac{\partial}{\partial h}\Delta D\frac{\partial}{\partial h}+\frac{\partial}{\partial h}\Delta\omega h\bigg)P_0 \nonumber\, ,
\end{eqnarray}
where G is the Green function defined in (\ref{76}) below. To
evaluate $P_1$, $P_2$, $\cdots$ we are repeatedly faced with the
problem of finding J, where
\begin{equation}
\int G k_{abc\dots}h_{a}'h_{b}'h_{c}'\dots P_0'\Pi\; dh'=J\left(h\right)
\label{75}
\end{equation}
or since symbolically
\begin{equation}
\sum\frac{\partial}{\partial
  h}\left(D\frac{\partial}{\partial h}+\omega h\right)G=\Pi\delta
\label{76}
\end{equation}
and defining $\widetilde{G}$ by
\begin{multline}
Khhh\dots P_0=\frac{\partial}{\partial
  h}\left(D\frac{\partial}{\partial h}+\omega h\right)JP_0\\
=P_0\left(D\frac{\partial}{\partial h}-\omega h\right)\left(-\frac{\partial}{\partial h}\right)J=P_0\widetilde{G}J
\label{77}
\end{multline}
we will show that the significant problem (significant to order $V^{-1}$
relative to other terms), is when $K_{abc \dots}h_a h_b h_c \dots $ has
none of $a,\, b,\, c,\, \dots$ paired {\it i.e.} $b,\, c\, \dots \neq
-a$. Then we try $J={\cal J}_{abc} h_a h_b h_c \dots$ we find a
solution provided
\begin{equation}
{\cal J}_{abc\dots }=\omega_a+\omega_b+\omega_c+\dots
\label{78}
\end{equation}\\
by direct substitution. The second derivative always gives zero, and
\begin{equation}
\omega_a h_a \frac{\partial}{\partial h_a}h_a=\omega_a h_a
\label{79}
\end{equation}
If ever we do find $h_{\bf l} h_{-l}$ we replace it by $\phi_{\bf l}+\left(h_{\bf l}
  h_{-l}-\phi_{\bf l} \right)$ and the bracketed term, as in all field theories
leads only to terms of order $V^{-1}$. Thus as the series
develops, a term in $h$ adds an $h$, $a\frac{\partial}{\partial
h}$ removes an $h$, a $G$ inserts a term like $\sum\omega$ for
each $h$. The first correction to $P_0$, $P_1$, illustrates the
process:
\begin{eqnarray}
P_1=\sum GM_{{\bf{{\bf kjl}}}}h_{{\bf{j}}}h_{{\bf{l}}}\frac{\partial}{\partial h_{{\bf{k}}}}P_0&\nonumber\\
&=&-\sum \left[GM_{{\bf{{\bf kjl}}}}h_{{\bf{j}}}h_{{\bf{l}}}h_{{\bf{-k}}}/\phi_{{\bf{k}}}\right]P_0\label{714}\\
&=&-\sum
P_0\tilde{G}M_{{\bf{{\bf kjl}}}}h_{{\bf{j}}}h_{{\bf{l}}}h_{{\bf{-k}}}/\phi_{{\bf{k}}}\label{715}\\
&=&-\sum
P_0\left(\frac{M_{{\bf{{\bf kjl}}}}h_{{\bf{-k}}}h_{{\bf{j}}}h_{{\bf{l}}}/\phi_{{\bf{k}}}}{\omega_{{\bf{k}}}+\omega_{{\bf{j}}}+\omega_{{\bf{l}}}}\right)\label{716}.
\end{eqnarray}

Suppose that we wanted the value of $\left<h_ah_bh_c\right>$, then
it is given to order $M$ by
\begin{equation}
-\frac{M_{{\bf{{\bf kjl}}}}\phi_{{\bf{j}}}\phi_{{\bf{l}}}}{\omega_{{\bf{k}}}+\omega_{{\bf{j}}}+\omega_{{\bf{l}}}}\left[\delta_{\bf{k},-\bf{a}}\delta_{\bf{j},-\bf{b}}+\mbox{permutations}\right].
\label{717}
\end{equation}
(Note that $M$ has $\delta_{\bf{k},\bf{j}+\bf{l}}$) At this point
we introduce the diagrams. Draw these pictures:

\begin{eqnarray}
\frac{\partial}{\partial
h_{{\bf{k}}}}Mh_{{\bf{j}}}h_{{\bf{l}}}&=&\quad
\raisebox{-8ex}{\includegraphics[height=16ex]{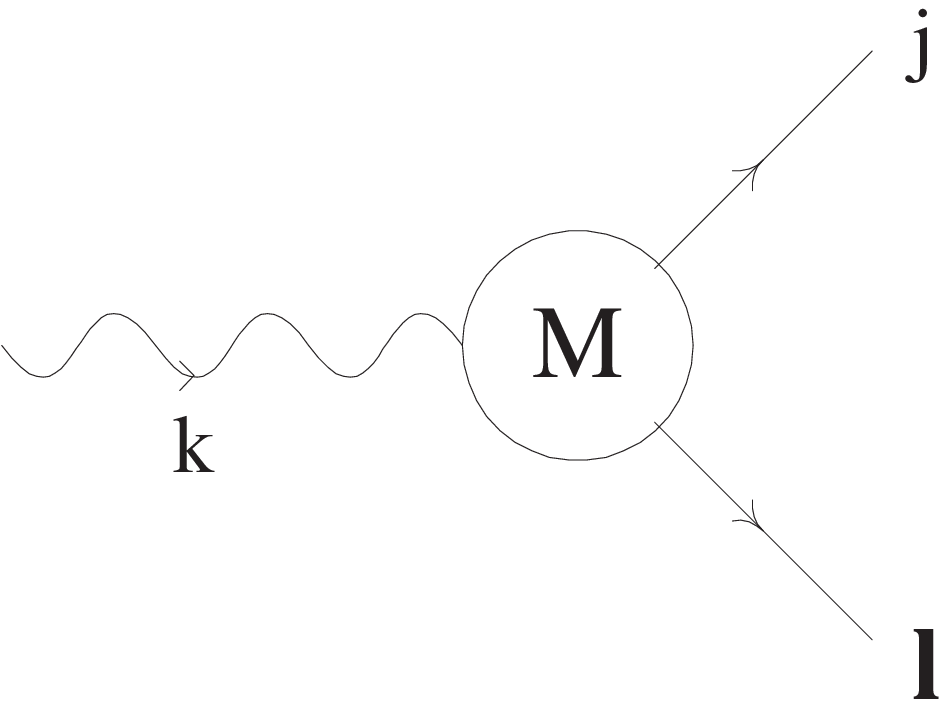}}
\end{eqnarray}

\parbox{11cm}{
\begin{eqnarray*}
\frac{\partial}{\partial h_{\bf k}}\frac{\partial}{\partial
h_{-k}}D_0&=&\quad
\raisebox{-6ex}{\includegraphics[height=12ex]{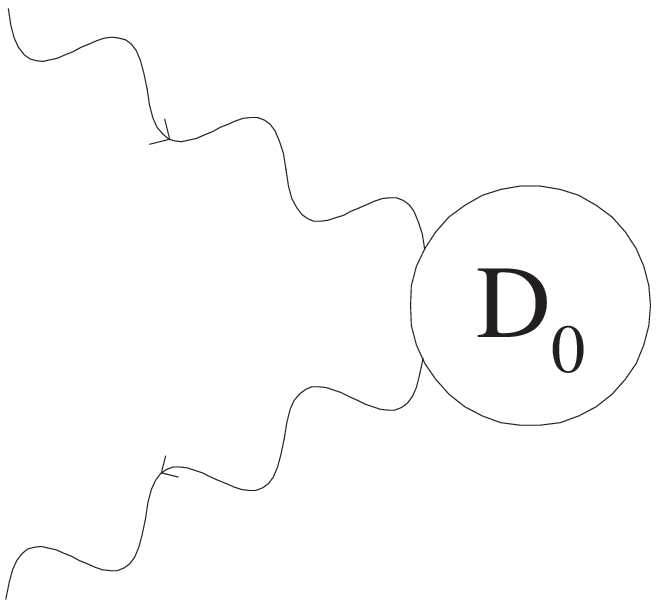}}\\
\frac{\partial}{\partial h_{\bf k}}\frac{\partial}{\partial
h_{-k}}\Delta D&=&\quad
\raisebox{-6ex}{\includegraphics[height=12ex]{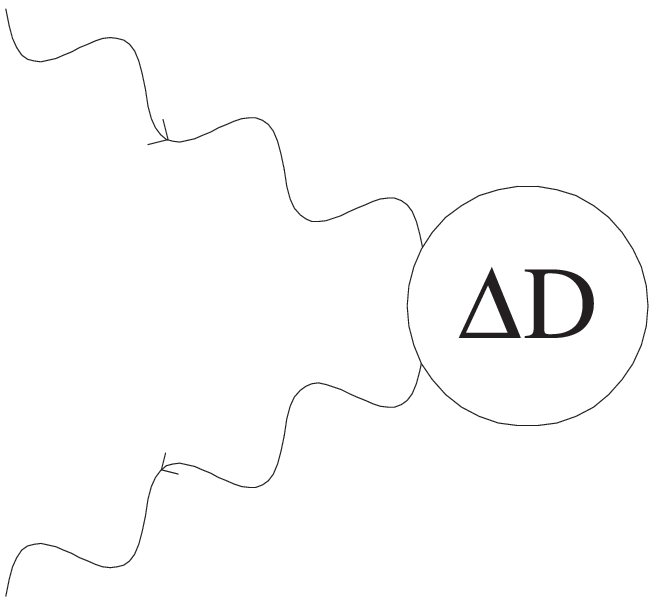}}
\end{eqnarray*}}
\parbox{1cm}{\begin{eqnarray}\end{eqnarray}\label{719}}

\parbox{11cm}{
\begin{eqnarray*}
\frac{\partial}{\partial h_{\bf k}}\nu_{\bf k}h_{\bf k}&=&\quad
\raisebox{-6ex}{\includegraphics[height=12ex]{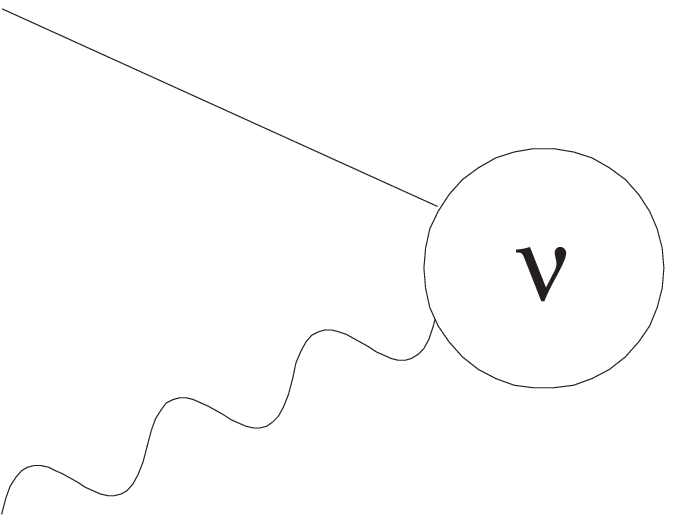}}\\
\frac{\partial}{\partial h_{\bf k}}h_{\bf k}\Delta\omega&=&\quad
\raisebox{-6ex}{\includegraphics[height=12ex]{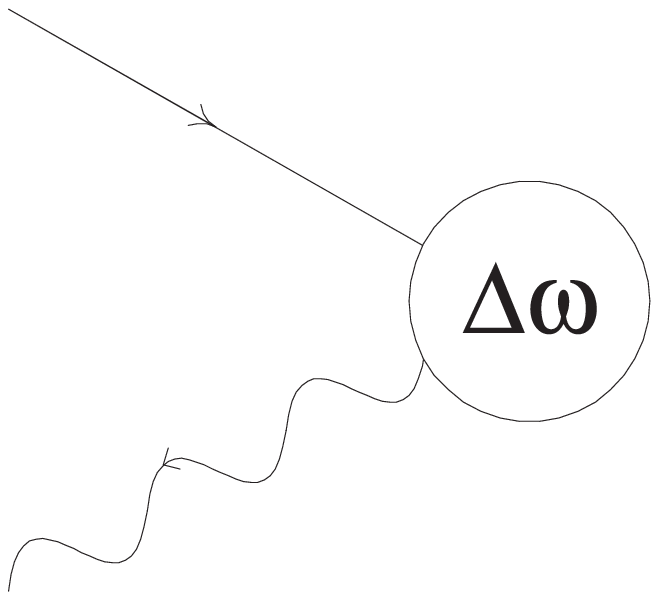}}
\end{eqnarray*}}
\parbox{1cm}{\begin{eqnarray}\end{eqnarray}\label{720}}

The rippled line will always point in the direction to the left,
for in final evaluations integration by parts gives non-zero
values only when the $\frac{\partial}{\partial h}$ finds an $h$ to
its left. The full lines representing $h$ can point left or right.
Note that $\omega_{{\bf k}}=\omega_{-{\bf k}}$ and that as in
normal field theories, strictly speaking arrows need to be
attached to the lines.

\centerline{\raisebox{-5ex}{\includegraphics[height=6ex]{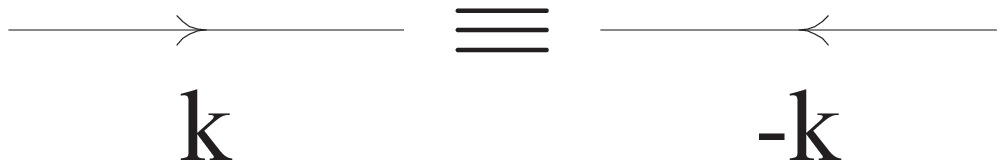}}}

Our problem now is:
\begin{equation}
\left( \raisebox{-4.5ex}{\includegraphics[height=10ex]{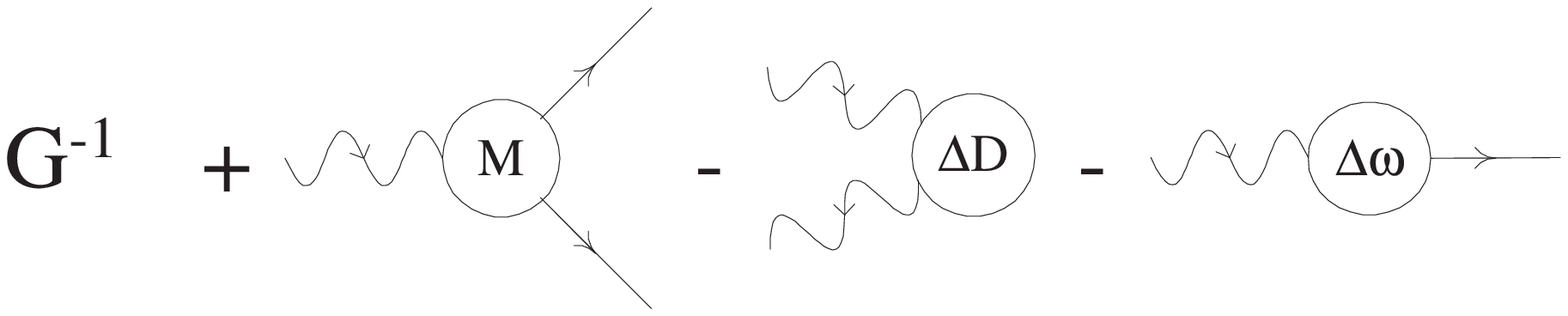}}
\right)P=0 \label{721}
\end{equation}

Upon integration, $P_0$ joins $\bf k$ and $- \bf k$ to give the
line

\begin{equation}
\raisebox{-3.3ex}{\includegraphics[height=5ex]{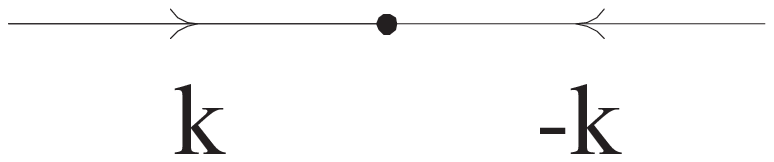}} =\phi_{\bf k}
\label{722}
\end{equation}

Thus our series for $P$ is
\begin{multline}
P=\Bigg( \raisebox{-10ex}{\includegraphics[height=20ex]{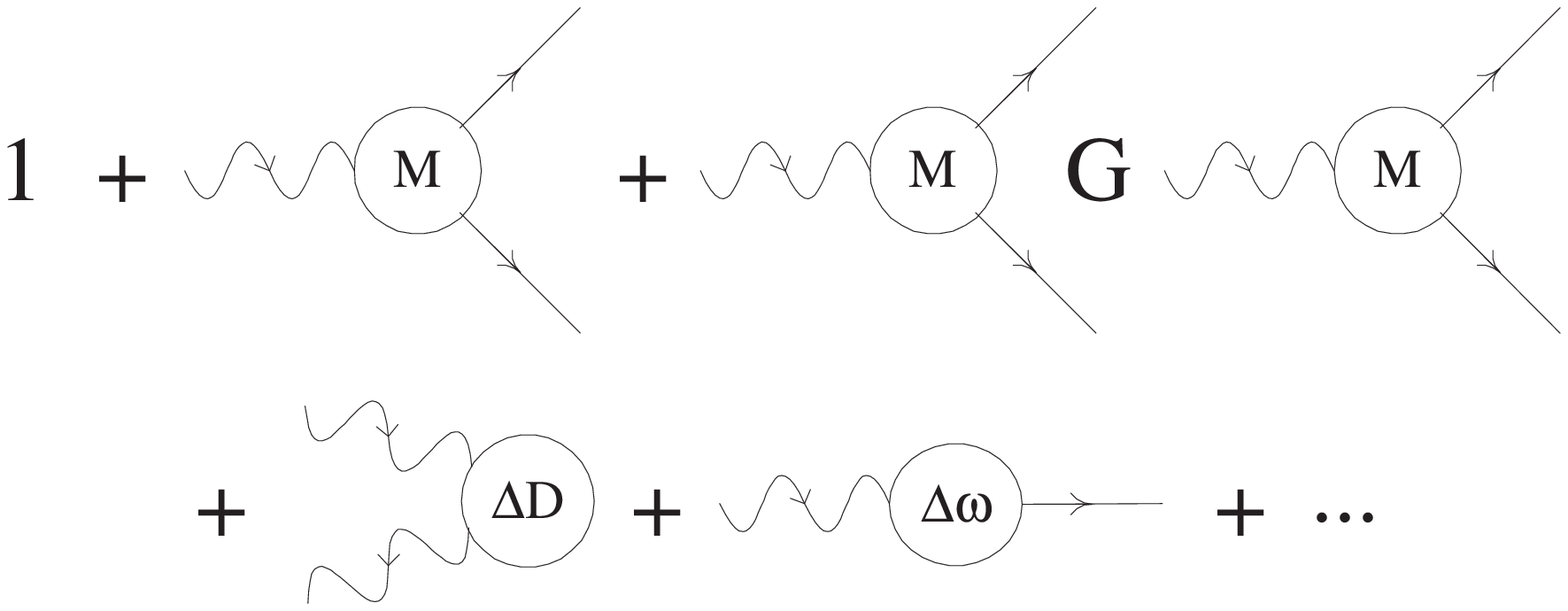}}
\Bigg)P_0 \label{723}
\end{multline}

The non linear coupling $M_{{\bf kjl}}$ used so far depend on the size
of the system. In fact $M_{{\bf kjl}}=\frac{1}{\sqrt{V}}m_{{\bf kjl}}$ where
$m_{{\bf kjl}}$ is of order 1. In the integrals appearing in the
following we redefine the $M$'s to be identical to the $m$'s.
Namely, the $M$'s that appear in the following are of order 1.

The condition $\phi_{\bf k} =\left<h_{\bf k}h_{-k}\right>$ as expressed in eq.
(\ref {114}) becomes

\begin{equation}
\raisebox{-25ex}{\includegraphics[height=50ex]{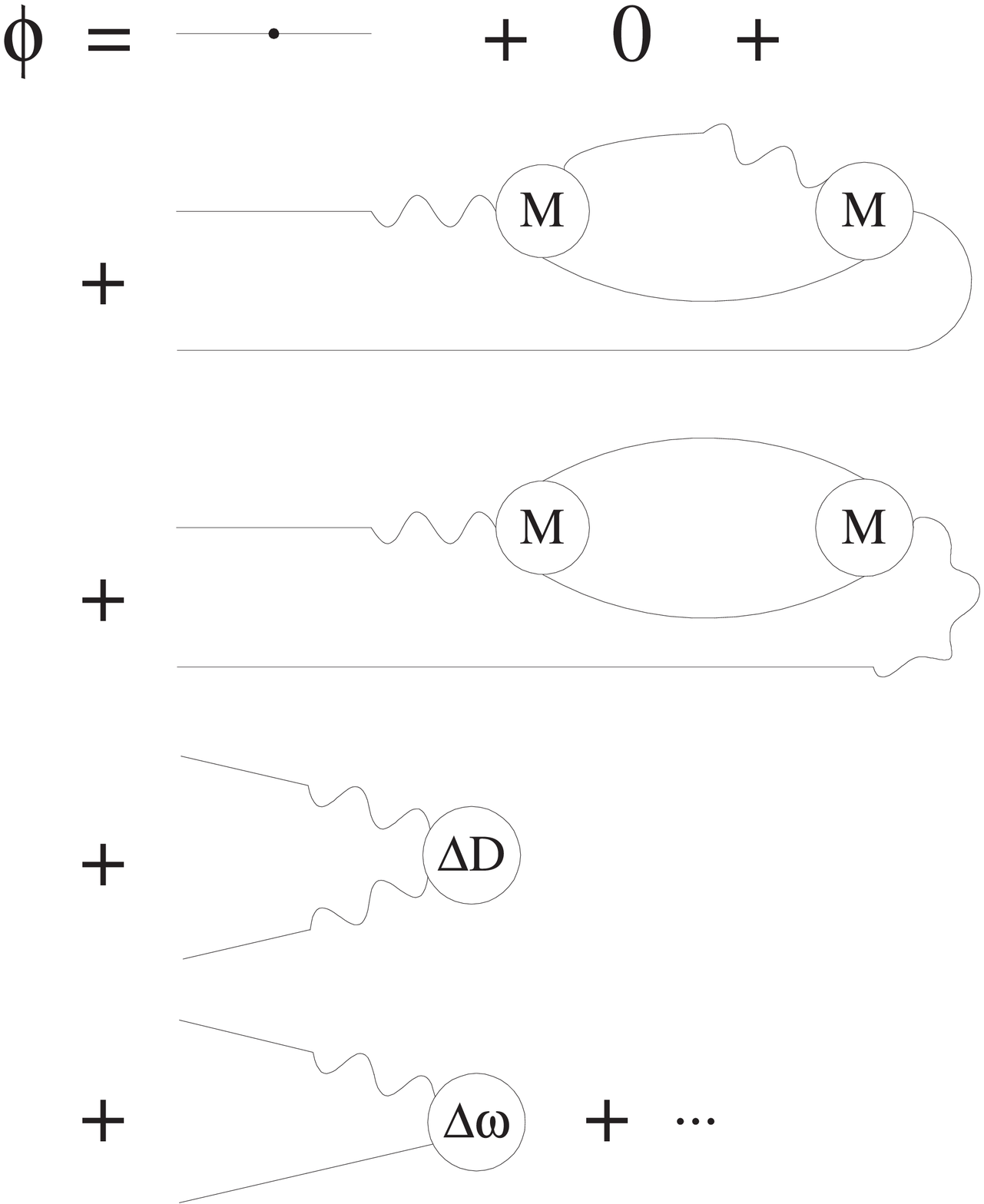}}
\label{725}
\end{equation}
\\to which we make this commentary: $\phi_{\bf k}=\phi_{\bf k}$ by definition, 0 comes from $<Mhhh>$ for odd
averages must be zero, or alternatively there is no way we can
join up two lines with
\raisebox{-6ex}{\includegraphics[height=12ex]{7_18.eps}}
.\\
The next diagram\\
\raisebox{-5.5ex}{\includegraphics[height=11ex]{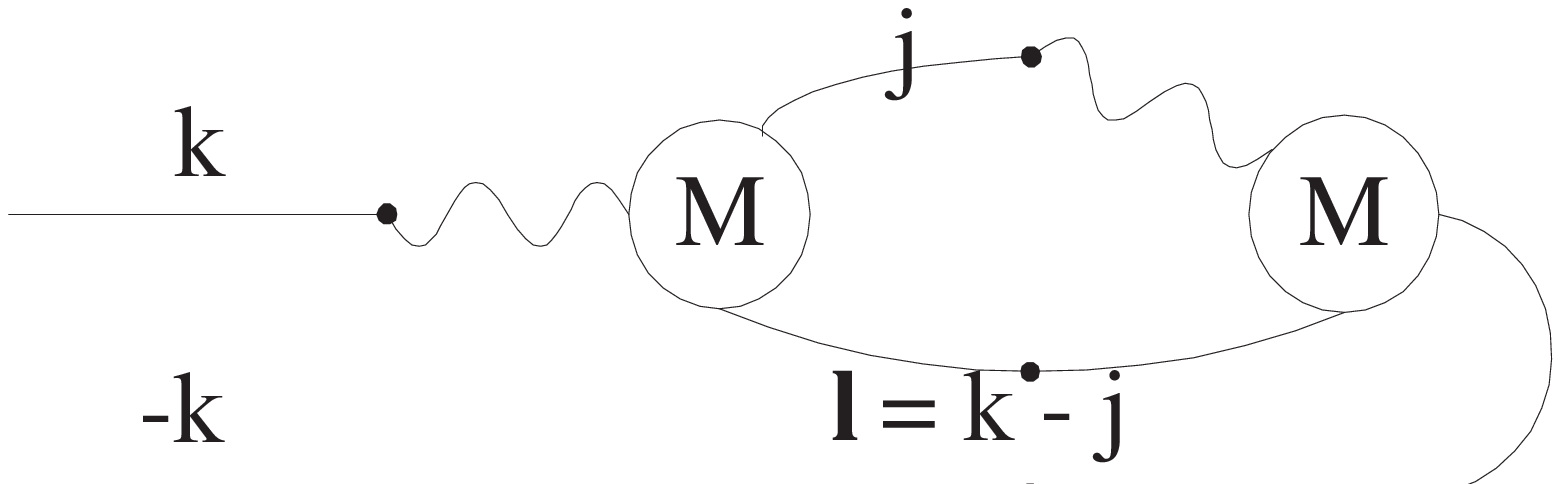}}\\
represents
\begin{equation}
\phi_{\bf k}
\int\frac{1M_{{\bf kjl}}1\phi_{\bf l}M_{\bf{jk},-\bf{l}}}{\omega_{\bf k}+\omega_{\bf j}+\omega_{\bf l}}d^3{\bf
j}, \label{726}
\end{equation}
(In the integrals we write for simplicity of presentation $l$,
meaning $l=k-j$ as shown in the diagram. This is the convention we
adopt in the following) where 1 (unity) is the result of
\begin{fmffile}{alan23}
\parbox{15mm}{\begin{fmfgraph*}(20,15)
\fmfleft{i1} \fmfright{o1} \fmf{plain}{i1,v1} \fmf{wiggly}{v1,o1}
\fmfv{d.sh=circle,d.f=1,d.size=2thin}{v1}
\end{fmfgraph*}}
\end{fmffile}
\hspace{3ex} i.e. $h\frac{\partial}{\partial h}$ when integrated
by parts, and $\left(\sum\omega\right)^{-1}$ comes from the Green
function as in (\ref{78}). A useful way to find this factor is to
put a vertical line at each point where a Green function
originally lay and put an $\omega$ for each line crossed including
these lines that yield unity: For example in the diagram above,
\\[2ex]
\centerline{\raisebox{-6ex}{\includegraphics[height=11ex]{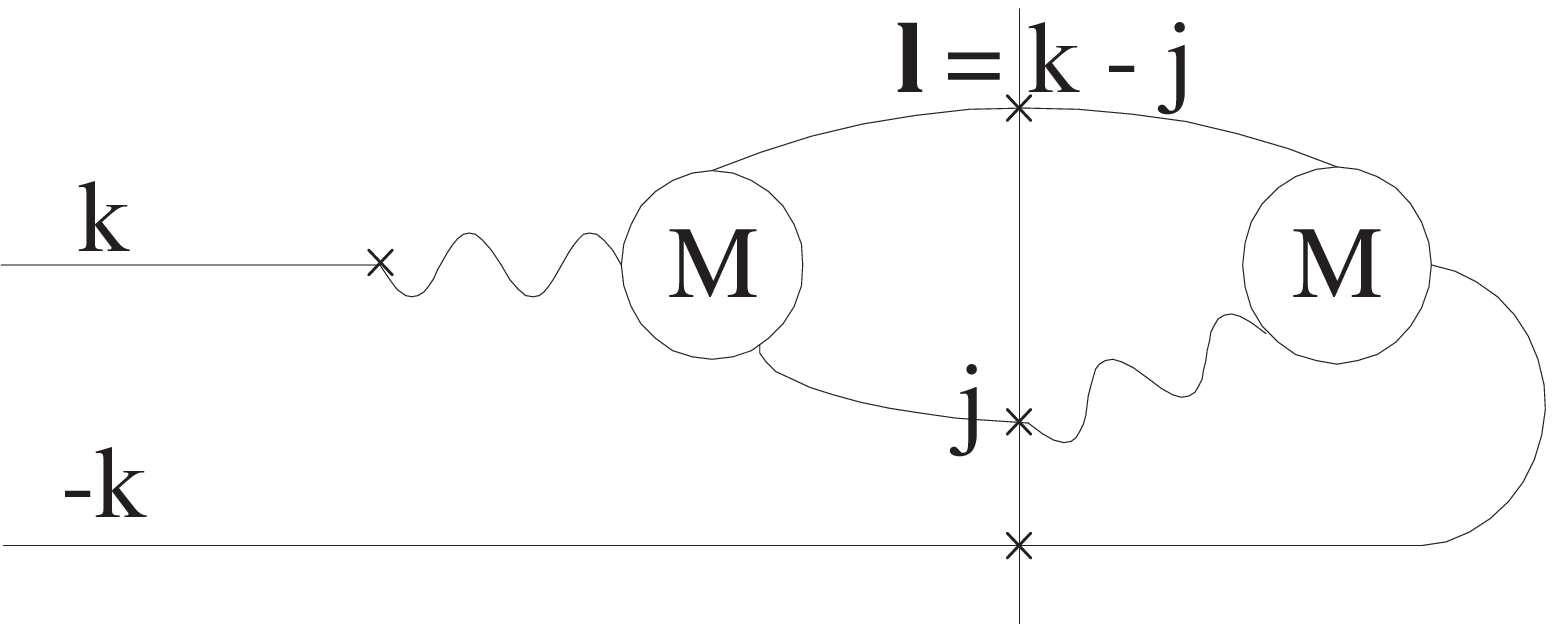}}}
we obtain a factor of
$\left(\omega_{\bf k}+\omega_{\bf j}+\omega_{\bf l}\right)^{-1}$.

Likewise,
\\[2ex]
\centerline{\raisebox{-6ex}{\includegraphics[height=11ex]{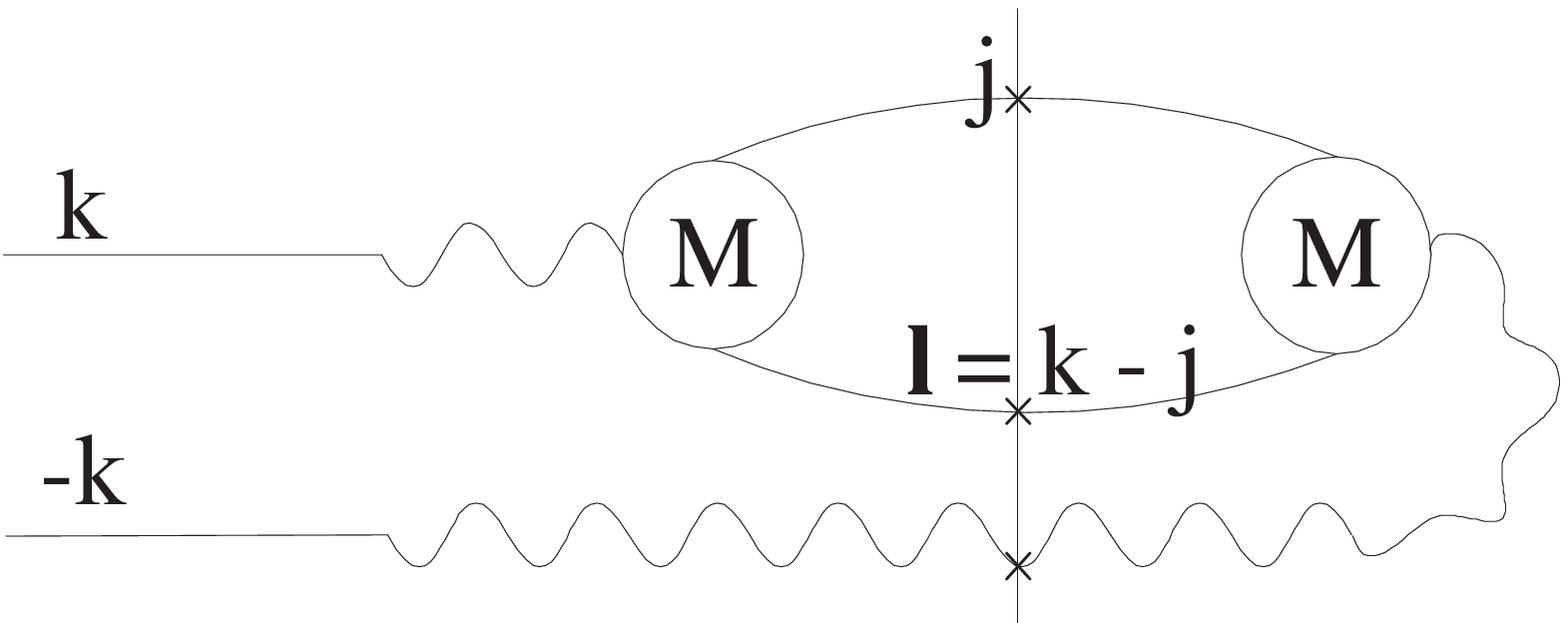}}}
represents
\begin{equation}
\int\frac{M_{{\bf kjl}}M_{-k,-j,-l}\phi_{\bf j}\phi_{\bf l}}{\omega_{\bf k}+\omega_{\bf j}+\omega_{\bf l}}d^3{\bf
j}. \label{728}
\end{equation}
The representation (\ref{725}) gives the steady transport equation
(\ref{114}). We now proceed to give a brief account of higher
orders. Firstly we can continue (\ref{114}) to the next order,
which must be $M^4$ as $M^3$ will give zero. There are as many
terms, so we just give typical terms. The four M's give

\begin{fmffile}{alan24i}
\parbox{25mm}{\begin{fmfgraph*}(20,15)
\fmfleft{i1} \fmfstraight \fmfrightn{o}{3} \fmf{wiggly}{i1,o2}
\fmf{plain}{o1,o2} \fmf{plain}{o3,o2}
\fmfv{d.sh=circle,d.f=empty,d.size=.5w,l=$M$,label.dist=0}{o2}
\end{fmfgraph*}}
\end{fmffile}
G
\begin{fmffile}{alan24i}
\parbox{25mm}{\begin{fmfgraph*}(20,15)
\fmfleft{i1} \fmfstraight \fmfrightn{o}{3} \fmf{wiggly}{i1,o2}
\fmf{plain}{o1,o2} \fmf{plain}{o3,o2}
\fmfv{d.sh=circle,d.f=empty,d.size=.5w,l=$M$,label.dist=0}{o2}
\end{fmfgraph*}}
\end{fmffile}
G
\begin{fmffile}{alan24i}
\parbox{25mm}{\begin{fmfgraph*}(20,15)
\fmfleft{i1} \fmfstraight \fmfrightn{o}{3} \fmf{wiggly}{i1,o2}
\fmf{plain}{o1,o2} \fmf{plain}{o3,o2}
\fmfv{d.sh=circle,d.f=empty,d.size=.5w,l=$M$,label.dist=0}{o2}
\end{fmfgraph*}}
\end{fmffile}
G
\begin{fmffile}{alan24i}
\parbox{25mm}{\begin{fmfgraph*}(20,15)
\fmfleft{i1} \fmfstraight \fmfrightn{o}{3} \fmf{wiggly}{i1,o2}
\fmf{plain}{o1,o2} \fmf{plain}{o3,o2}
\fmfv{d.sh=circle,d.f=empty,d.size=.5w,l=$M$,label.dist=0}{o2}
\end{fmfgraph*}}
\end{fmffile}\\
and two typical terms are

\vspace{3ex}
\centerline{\raisebox{-6ex}{\includegraphics[height=12ex]{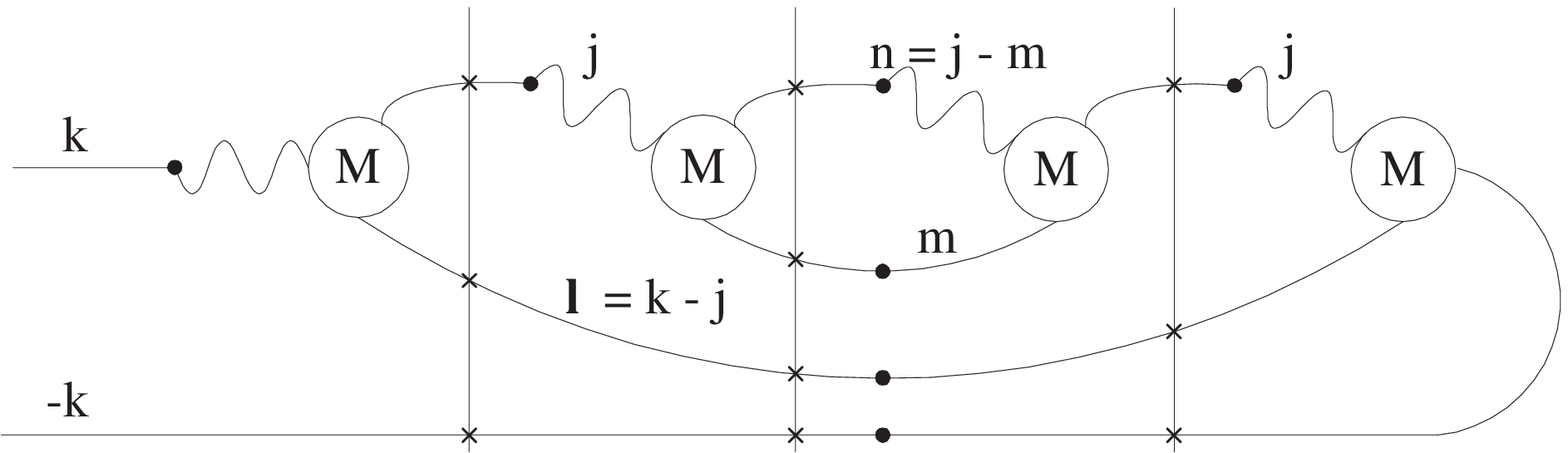}}}
\vspace{3ex}
\begin{equation}
=\int\frac{M_{{\bf kjl}}M_{jmn}M_{nj,-m}M_{\bf{jk},-\bf{l}}\phi_{\bf l}\phi_{\bf m}\phi_{\bf k}}{(\omega_{\bf j}+\omega_{\bf l}+\omega_{\bf k})(\omega_{\bf l}+\omega_{\bf m}+\omega_{\bf n}+\omega_{\bf k})(\omega_{\bf j}+\omega_{\bf l}+
  \omega_{\bf k})} {\rm d}^3{\bf l}\, {\rm d}^3{\bf m}
\label{729}
\end{equation}

\vspace{3ex}
\centerline{\raisebox{-6ex}{\includegraphics[height=12ex]{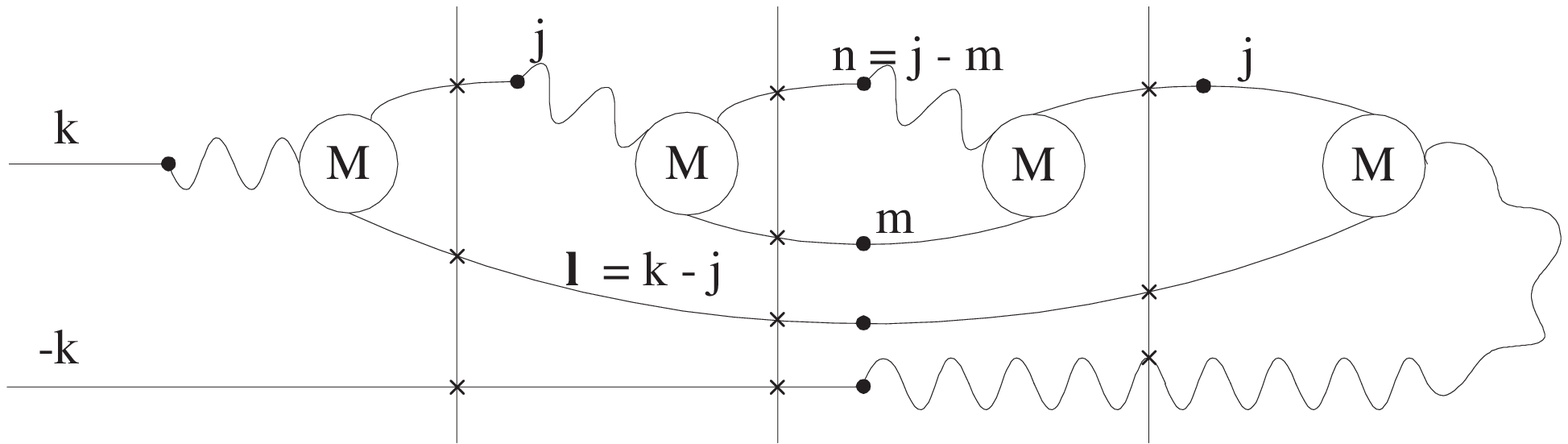}}}
\vspace{3ex}

\begin{equation}
=\int
\frac{M_{{\bf kjl}}M_{jmn}M_{nj,-m}M_{-k,-j,-l}\phi_{\bf l}\phi_{\bf m}\phi_{\bf j}}{(\omega_{\bf j}+\omega_{\bf l}+\omega_{\bf k})(\omega_{\bf n}+\omega_{\bf m}+\omega_{\bf l}+\omega_{\bf k})(\omega_{\bf j}+\omega_{\bf l}+\omega_{\bf k})}
 {\rm d}^3{\bf l}\, {\rm d}^3{\bf m}
\label{730}
\end{equation}
and cross terms with $\Delta D$ and $\Delta\omega$ e.g.

\vspace{3ex}
\centerline{\raisebox{-6ex}{\includegraphics[height=12ex]{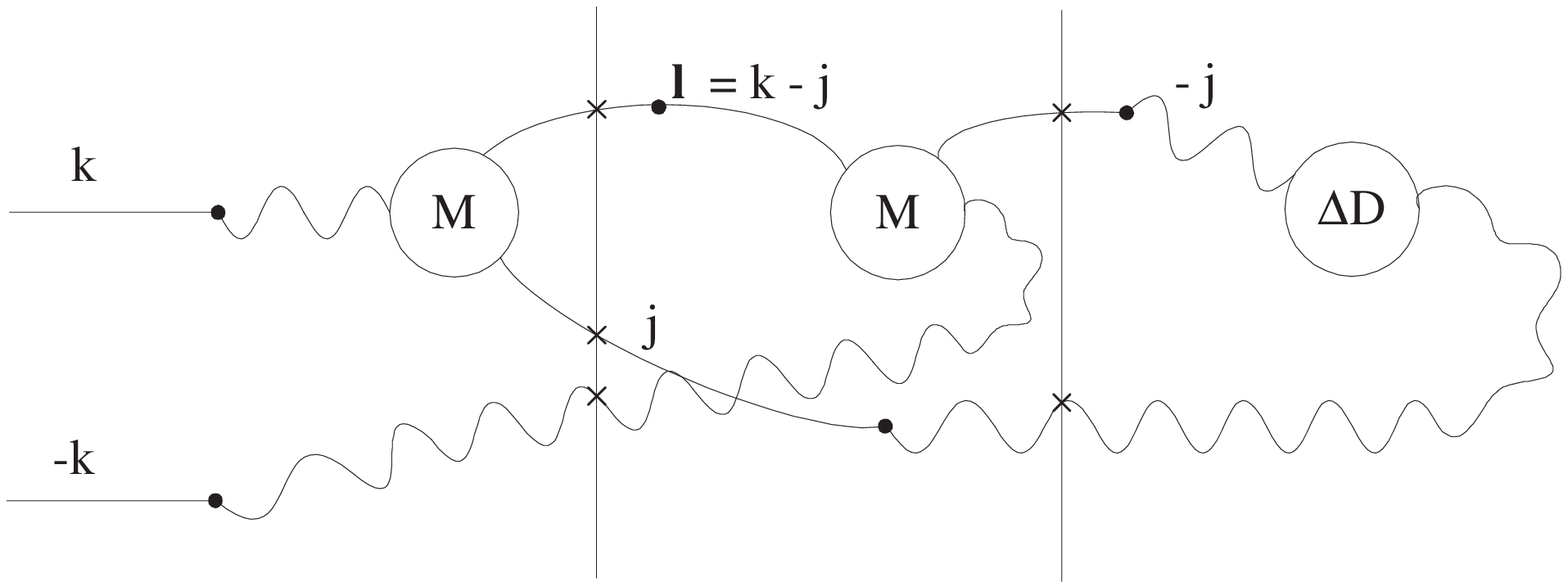}}}
\vspace{3ex}

\begin{equation}
=\int \frac{M_{{\bf kjl}}M_{-k,-j,-l}\phi_{\bf l}\phi_{\bf j} (\Delta
  D)_{\bf j}}{(\omega_{\bf j}+\omega_{\bf l}+\omega_{\bf k})2\omega_{\bf j}}\; d^3{\bf j}
\label{731}
\end{equation}
($\omega_{\bf k}=\omega_{-k}$).

Next we consider the value of $\left<h_{\bf k}h_{-j}h_{-l}\right>$ which
has the first approximation as above in (\ref{717}).

\vspace{3ex} \centerline{$-\Bigg\{$
  {\raisebox{-6ex}{\includegraphics[height=12ex]{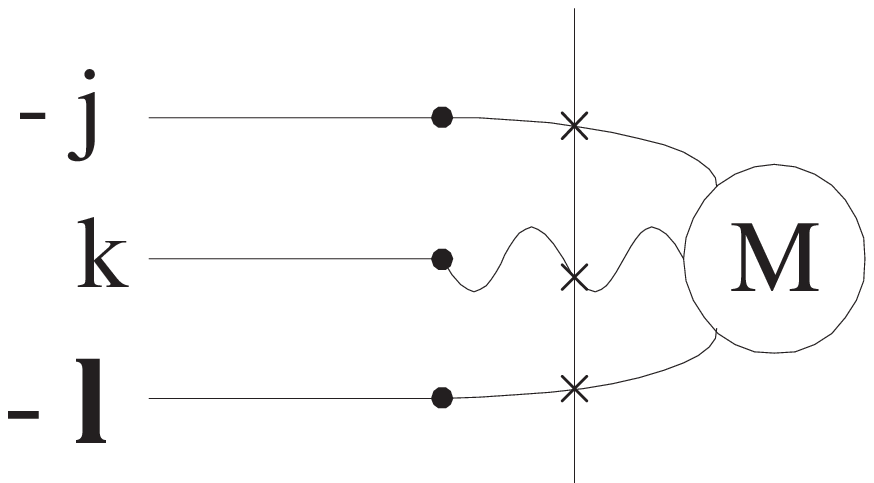}}} +
  permutations $\Bigg\}$}
\vspace{3ex}

\begin{equation}
=-\frac{1}{\sqrt{V}}\frac{M_{{\bf kjl}}\phi_{\bf j}\phi_{\bf l}}{\omega_{\bf k}+\omega_{\bf j}+\omega_{\bf l}}+\hbox{permutations}
\label{732}
\end{equation}

Corrections appear at order $M^3$ typically

\vspace{3ex}
\centerline{\raisebox{-6ex}{\includegraphics[height=12ex]{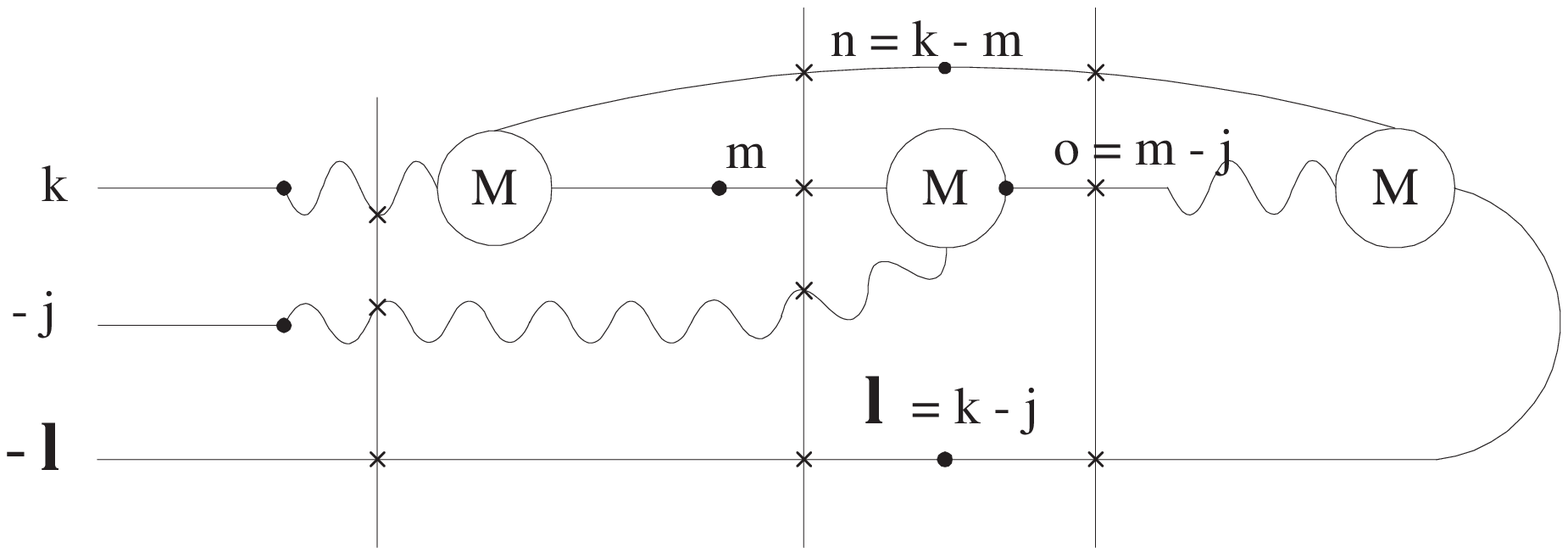}}}
\vspace{3ex}

\begin{equation}
=\frac{1}{\sqrt{V}} \int
\frac{M_{kmn}M_{-j,o,-m}M_{o,l,-n}\phi_{\bf l}\phi_{\bf m}\phi_{\bf n}}{(\omega_{\bf j}+\omega_{\bf l}+\omega_{\bf k})(\omega_{\bf l}+\omega_{\bf n}+\omega_o)(\omega_{\bf j}+\omega_{\bf l}+\omega_{\bf m}+\omega_{\bf n})}\;
d^3{\bf m} \label{733}
\end{equation}

Finally the four h correlation is given by

\begin{eqnarray*}
& &\raisebox{-5ex}{\includegraphics[height=10ex]{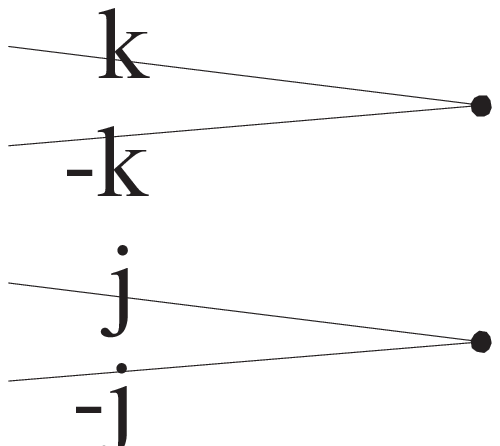}} \hbox{
+
  permutations}\\
&+& \raisebox{-6ex}{\includegraphics[height=12ex]{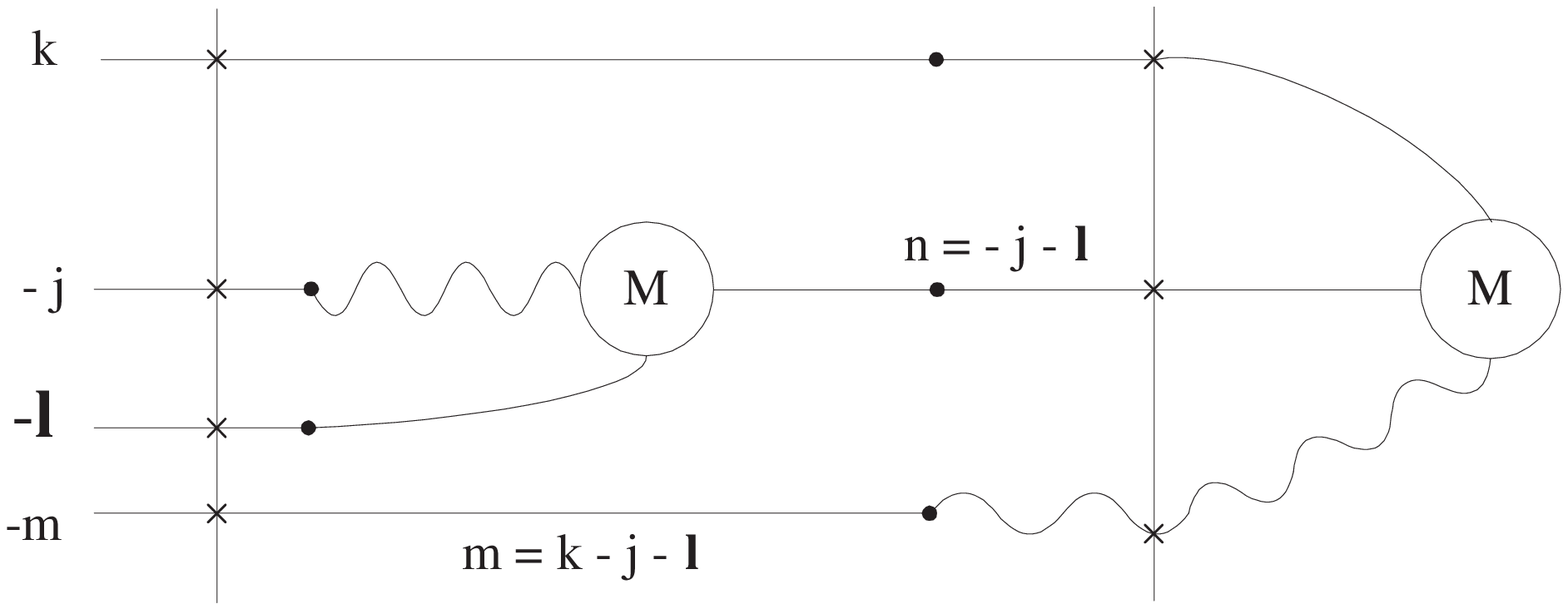}}\\
&+& \raisebox{-6ex}{\includegraphics[height=12ex]{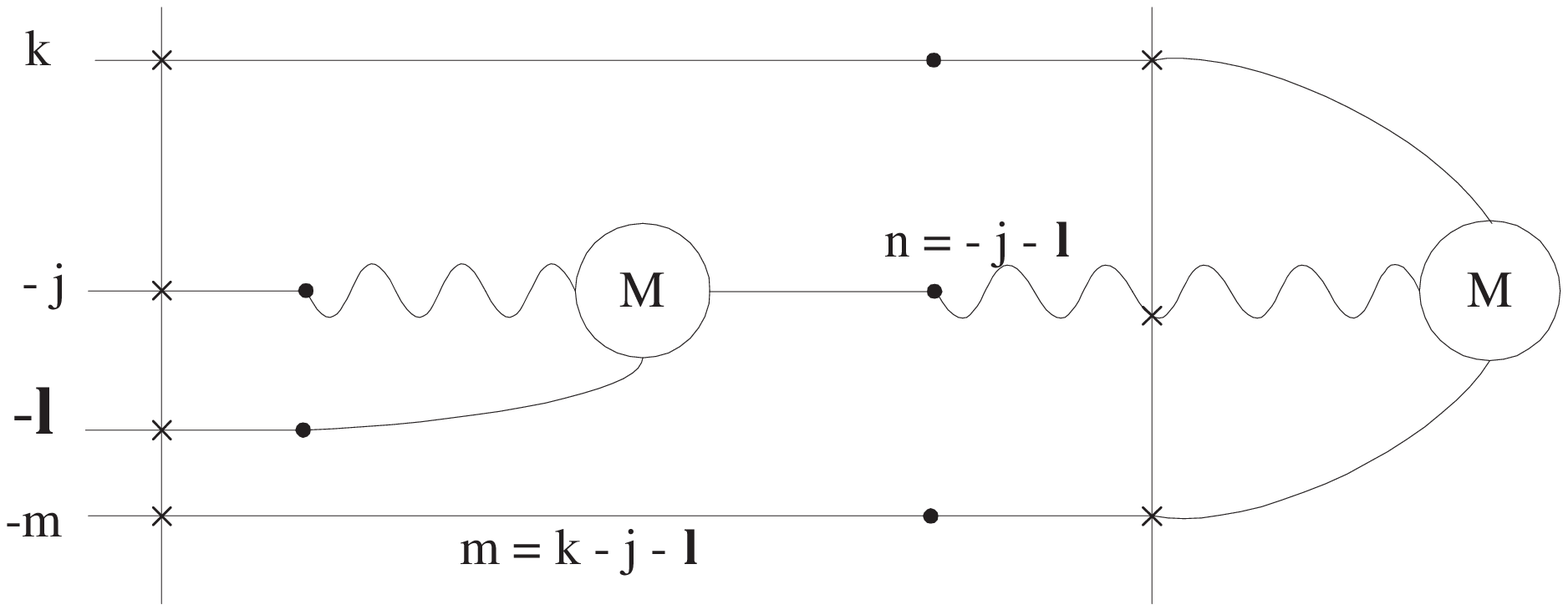}}\;
+\cdots
\end{eqnarray*}

\begin{multline}
\left<h_{\bf k} h_{-j} h_{-l} h_{-m} \right>=\phi_{\bf k} \phi_{\bf j}\delta_{k,l}\delta_{j,-m}+\phi_{\bf k} \phi_{\bf j}\delta_{k,m}\delta_{j,-l}+\phi_{\bf k} \phi_{\bf l}\delta_{k,j}\delta_{l,-m}+\\
+\frac{1}{V}\frac{M_{-j,l,n}M_{-m,-n,-k}\phi_{\bf k}\phi_{\bf l}\phi_{\bf n}}{(\omega_{\bf k}+\omega_{\bf j}+\omega_{\bf l}+\omega_{\bf m})(\omega_{\bf k}+\omega_{\bf m}+\omega_{\bf n})}\\
+\frac{1}{V}\frac{M_{-j,l,n}M_{n,-k,m}\phi_{\bf k}\phi_{\bf l}\phi_{\bf m}}{(\omega_{\bf k}+\omega_{\bf j}+\omega_{\bf l}+\omega_{\bf m})(\omega_{\bf k}+\omega_{\bf m}+\omega_{\bf n})}
+ \cdots \label{735}
\end{multline}

and many similar terms, multiplying vastly at the next $M^4$
order. Notice that the series can be thought of as an expansion in
\begin{eqnarray*}
\frac{M^2\phi}{\omega^2}k^d
\end{eqnarray*}
and as has been noted earlier, a scaling relationship essential
for the plausibility of the expansion, requires this combination
to have a behaviour like $k^0$. For the time dependent equations
the graphology holds with the 3 vector $\bf{k}$ replaced by a
four vector, but with one other change which is the interference
of the external noise with the non linear terms, however this does
not have an effect until a higher order is reached than the second
order we have used in this paper, and so we will not discuss it
further.

\newpage
\section{}
\setcounter{equation}{0}
 At first sight it is
unusual for an integral operator to give $\delta$ functions, but
this can be illustrated by a simple model. Consider:
\begin{equation}
\int_0^{\infty}\frac{dy}{\left(x+y\right)^2}\sqrt{\frac{y}{x}}-\int_0^{\infty}\frac{dy}{\left(x+y\right)^2}\sqrt{\frac{x}{y}}=Y\left(x\right).
\label{45}
\end{equation}
Clearly
\begin{equation}
\int_0^{\infty}Y\left(x\right)\;dx=0 \label{46}
\end{equation}
for one needs only to interchange $x$ and $y$ in either integral.
Also, we note that if $x\neq 0$, or $x\neq\infty$, then, by the
transformation $y=xz$:
\begin{equation}
\int_0^{\infty}\frac{dy}{\left(x+y\right)^2}\sqrt{\frac{y}{x}}=\frac{1}{x}\int_{0}^{\infty}\frac{dy}{\left(y+1\right)^2}\sqrt{y}
\label{47}
\end{equation}
which by the transformation $y\rightarrow 1/y$ equals
\begin{equation}
\frac{1}{x}\int_{0}^{\infty}\frac{dy}{\left(y+1\right)^2}\frac{1}{\sqrt{y}}
\label{48}
\end{equation}
which can now be returned to
\begin{equation}
\int_0^{\infty}\frac{dy}{\left(x+y\right)^2}\sqrt{\frac{x}{y}}.
\label{49}
\end{equation}
Hence if $x\neq 0$ then
\begin{equation}
\int_0^{\infty}\frac{dy}{\left(x+y\right)^2}\sqrt{\frac{y}{x}}=\int_0^{\infty}\frac{dy}{\left(x+y\right)^2}\sqrt{\frac{x}{y}}.
\label{410}
\end{equation}
However, if we break the integration of y from
\begin{equation}
\int_{0}^{\infty}\;\mbox{to}\;\int_{0}^{\epsilon}+\int_{\epsilon}^{\infty},
\label{411}
\end{equation}
we see that
\begin{multline}
\int_{0}^{\epsilon}\;dx\int_{0}^{\infty}\frac{dy}{\left(x+y\right)^2}\sqrt{\frac{y}{x}}-\int_{0}^{\epsilon}\;dx\int_{0}^{\infty}\frac{dy}{\left(x+y\right)^2}\sqrt{\frac{x}{y}}=\\\int_{0}^{\epsilon}\;dx\int_{0}^{\epsilon}\frac{dy}{\left(x+y\right)^2}\left(\sqrt{\frac{y}{x}}-\sqrt{\frac{x}{y}}\right)
+\int_{0}^{\epsilon}\;dx\int_{\epsilon}^{\infty}\frac{dy}{\left(x+y\right)^2}\left(\sqrt{\frac{y}{x}}-\sqrt{\frac{x}{y}}\right)
\label{412}
\end{multline}
The first term is zero by relabelling $x\rightleftarrows y$ in one
of the integrals, and the second term, by writing
$x\rightarrow\epsilon x$, and $y\rightarrow\epsilon y$
\begin{eqnarray}
=&\int_{0}^{1}\;dx\int_{1}^{\infty}\;dy\frac{1}{\left(x+y\right)^2}\frac{y-x}{\sqrt{xy}}\nonumber\\
=&\mbox{a constant $c>0$}\label{413}
\end{eqnarray}
since $y>x$ throughout the entire range.

So we have proved (using a similar proof at ${\bf{x}}=\infty$)
that
\begin{equation}
\int_0^{\infty}\frac{dy}{\left(x+y\right)^2}\sqrt{\frac{y}{x}}-\int_0^{\infty}\frac{dy}{\left(x+y\right)^2}\sqrt{\frac{x}{y}}=a\delta\left(x\right)-a\delta\left(x-\infty\right).
\label{414}
\end{equation}
This model can be rigorized in the manner of $\delta$ function
theory, but since we offer it simply to clarify the situation for
the reader we will not give the rigorous mathematics. The model is
a faithful representation of what goes on in equation (\ref{42}),
in this particular input, if short of the algebraic complexity of
the 3D Navier Stokes equation. This means that if we assume
Kolmogoroff is true and substitute $k^{-\Gamma}$ in equation
(\ref{114}) we can expect the non-linear terms to cancel at
$\Gamma~=~11/3$.

\newpage
\section{}
\setcounter{equation}{0} In this appendix we discuss in detail the
long time decay of $\Phi_{\bf{k}}(t)$, that follows from equation
(\ref{313}) To orient oneself, consider what the terms in
(\ref{313}) look like if the solution were an exponential decay.
The two "Boltzmann" terms decay like:

\parbox{11cm}{
\begin{eqnarray*}
&\Phi_{{\bf{k}}}\left(t\right)\;&\mbox{and}\;\Phi_{{\bf{k-j}}}\left(t\right)\Phi_{{\bf{j}}}\left(t\right)\nonumber\\
\mbox{i.e.}&
\rm{e}^{-\omega_{{\bf{k}}}t}\;&\mbox{and}\;\rm{e}^{-\left(\omega_{{\bf{k-j}}}+\omega_{{\bf{j}}}\right)t}.
\end{eqnarray*}}\hfill
\parbox{1cm}{\begin{eqnarray}\end{eqnarray}\label{51}}
Further, think of what would happen if $\omega_{{\bf{k}}}$ were
just viscosity, i.e. $k^2$. Then the locus of ${\bf{j}}$ when
\begin{equation}
k^2=\left({\bf{k}}-{\bf{j}}\right)^2+j^2 \label{52}
\end{equation}
is a $d$-dimensional sphere with ${\bf{k}}$ as its diameter, from
Pythagoras's theorem. For $\bf{j}$ inside the sphere
\begin{equation}
\omega_{{\bf{k}}}>\omega_{{\bf{k-j}}}+\omega_{{\bf{j}}} \label{53}
\end{equation}
and for $\bf{j}$ outside the sphere,
\begin{equation}
\omega_{{\bf{k}}}<\omega_{{\bf{k-j}}}+\omega_{{\bf{j}}}.
\label{54}
\end{equation}
It becomes now very clear that the contribution of $j$'s within
the sphere decays slower than $\Phi_{\bf k}(t)$ and cannot, therefore,
solve eq.(\ref{313}) for long times. The above discussion is
limited to decay rates that are proportional to $k^2$. It is well
known that in non-trivial cases the decay rates are proportional
to $k^{\mu}$, with $\mu\neq 2$, and in most cases (turbulence
being an exception) $\mu >1$. It is easy to verify that if $\mu
>1$ the minimum of $\omega_{\bf{k}-\bf{j}}+\omega_{\bf j}$ occurs at
$\bf{j}=\bf{k}/2$ and the value of $\omega_{\bf{k}-\bf{j}}+\omega_{\bf j}$ at
the minimum is $2^{1-\mu}\omega_{\bf k}$. The only difference from the
case of $\mu =2$ is the shape of the region where
$\omega_{\bf{k}-\bf{j}}+\omega_{\bf j}<\omega_{\bf k}$ in which the vector $\bf{k}$
plays the role of an axis of symmetry and its end points are on
the boundary of the region. The conclusion is that the single mode
decay that can be a quite reasonable description of $\Phi_{\bf k}(t)$
for relatively short times $(\omega_{\bf k}t\leq 1)$ cannot describe the
long time decay. Since the argument depends on whether $\mu$ is
larger or smaller than 1, we will discuss in the following the two
cases separately.

\textbf{a}. The situation $\mu >1$ is generic, it appears in
$KPZ$, dynamics at the transition of the $\phi^4$ theory etc. We
expect that, for long times,
\begin{equation}
\Phi_{\bf k}(t)=\sum_{i=1}^{\infty}\psi_i(k)f_i(\omega_{\bf k}t) \label{55}
\end{equation}
such that
\begin{equation}
\lim_{x\to\infty}\frac{|f_{m+1}(x)|}{|f_{\bf m}(x)|}=0 \label{56}
\end{equation}
and the $f_i(0)$'s are chosen to be 1. If the series in (\ref{55})
converges for $t=0$, then
\begin{equation}
\sum_{i=1}^{\infty}\psi_i(k)=\phi(k) \label{57}
\end{equation}
but there is no reason to assume a scaling function of the form
$\phi_qf(\omega_qt)$. In fact, since the cases we consider have
$D_{{\bf k}0}\neq 0$ for $k\neq 0$, this is actually impossible, because
of the exact relation (\ref{317}) that is also obeyed by our
approximation
\begin{equation}
\lim_{t\to 0+}\frac{\partial \Phi_{\bf k}(t)}{\partial t}=-_{{\bf k}0}.
\end{equation}
This is possible with the single scaling function form only if
$\omega_{\bf k}=\frac{_{{\bf k}0}}{f'(0)\phi_{\bf k}}$. This is not normally the
case, $\omega_{\bf k}$ does not scale as $_{{\bf k}0}/\phi_{\bf k}$. Our aim is to
obtain the functional form of $f_1(x)$. We assume first that the
non-linear term is not relevant at large $t$, and arrive at a
contradiction. If it is not relevant, eq.(\ref{313}) gives
\begin{equation}
\frac{\partial \Phi_{\bf k}(t)}{\partial
t}+\left\{\nu_{\bf k}+\int\Lambda^{(1)}(j,k)\phi_{\bf j}\right\}\Phi_{\bf k}=0
\label{58}
\end{equation}
and the solution is
\begin{equation}
\Phi_{\bf k}(t)=\phi_{\bf k}{\rm e}^{-\omega_{\bf k}t} \label{59}
\end{equation}
with
\begin{equation}
\omega_{\bf k}=\nu_{\bf k}+\int\Lambda^{(1)}(j,k)\phi_{\bf j}\;{\rm d}j \label{510}
\end{equation}
but since the $\omega_{\bf k}$ obtained here behaves like $k^{\mu}$ for
small $\mu$, with $\mu >1$, such a solution is not consistent with
the assumption that the non-linear term is not relevant. By the
same reasoning, leading to the conclusion the ${\rm
e}^{-\omega_{\bf k}t}$ is not consistent as the description of the long
time decay, it becomes clear that any decay faster than an
exponential for $f(x)$ is not acceptable. For a decay that is
slower than an exponential the derivative can be neglected
compared to the function, for large $x$ and eq.(\ref{313}) may be
approximated for large $\omega_{\bf k}t$ by
\begin{equation}
\left\{\nu_{\bf k}+\int\Lambda^{(1)}(j,k)\phi_{\bf j}\;{\rm
d}j\right\}\Phi_{\bf k}(t)-\int\Lambda^{(2)}(j,k)\Phi_{\bf j}(t)\Phi_{\bf{k}-\bf{j}}(t)=0.
\label{511}
\end{equation}
Using the long time leading order behaviour we find
\begin{multline}
\left\{\nu_{\bf k}+\int\Lambda^{(1)}(j,k)\phi_{\bf j}\;{\rm
d}j\right\}\psi_{\bf k}^{(1)}f_1(\omega_{\bf k}t)\\-\int\Lambda^{(2)}(j,k)\psi_{\bf j}^{(1)}\psi_{\bf{k}-\bf{j}}^{(1)}f_1(\omega_{\bf j}t)f_1(\omega_{\bf{k}-\bf{j}}t)=0.
\label{512}
\end{multline}
Consider first the one dimensional case. We assume, without loss
of generality, that $k$ is positive and break up the integral on
the left hand side of eq.(\ref{512}) into two contributions,
\begin{multline}
\int_{-\infty}^{\infty}\Lambda^{(2)}\psi_{\bf j}^{(1)}\psi_{\bf{k}-\bf{j}}^{(1)}f_1(\omega_{\bf j}t)f_1(\omega_{\bf{k}-\bf{j}}t)\;{\rm d}j=\\
\int_0^k\Lambda^{(2)}(j,k)\psi_{\bf j}^{(1)}\psi_{\bf{k}-\bf{j}}^{(1)}f_1(\omega_{\bf j}t)f_1(\omega_{\bf{k}-\bf{j}}t)\;{\rm
d}j\\+\int_{\overline{[0,k]}}\Lambda^{(2)}(j,k)\psi_{\bf j}^{(1)}\psi_{\bf{k}-\bf{j}}^{(1)}f_1(\omega_{\bf j}t)f_1(\omega_{\bf{k}-\bf{j}}t)\;{\rm
d}j, \label{513}
\end{multline}
where $\overline{[0,k]}$ is the set complimentary to the segment
$[0,k]$. Because of the form of $\omega_{\bf k}$, it is obvious that the
second contribution on the right hand side of (\ref{513}) decays
faster than the first contribution. The leading order decay in
eq.(\ref{512}) will be thus obtained by equating
\begin{multline}
\left\{\nu_{\bf k}+\int\Lambda^{(1)}(j,k)\phi_{\bf j}\;{\rm d}j\right\}\psi_{\bf k}^{(1)}f_1(\omega_{\bf k}t)\\
=\int_0^k\Lambda^{(2)}(j,k)\psi_{\bf j}^{(1)}\psi_{\bf{k}-\bf{j}}^{(1)}f_1(\omega_{\bf j}t)f_1(\omega_{\bf{k}-\bf{j}}t)\;{\rm
d}j \label{515}
\end{multline}
Since this equation must hold for all long enough times it is
obvious that the only way to satisfy the equation is to have
$f(\omega_{\bf k}t)\propto{\rm e}^{-\gamma[\omega_{\bf k}t]^{1/\mu}}$, where
$\gamma$ is a numerical constant. The reason is that within the
range $0\leq j\leq k$
\begin{equation}
|\bf{k}|=|\bf{j}|+|\bf{k}-\bf{j}| \label{516}
\end{equation}
so that a single time decay can be brought out of the $j$ integral
and made equal to the decay of $\Phi_{\bf k}(t)$. (In fact, the
situation is a little more subtle than the above description,
because regardless how large is the time, within the range of
integration there are always $j$'s and (and $\bf{k}-\bf{j}$'s)
such that $\omega_{\bf j}t<1$ and there the long time form for the two
point function is unjustified. This region of small $j$'s
decreases however with $t$ and in the generic case where there are
no convergence difficulties at $j=0$ this becomes unimportant).
The coefficients $\psi_{\bf k}^{(1)}$ obey the equation
\begin{equation}
\left\{\nu_{\bf k}+\int\Lambda^{(1)}(j,k)\phi_{\bf j}\right\}\psi_{\bf k}^{(1)}-\int_0^{k}\Lambda^{(2)}(j,k)\psi_{\bf j}^{(1)}\psi_{{\bf j}-{\bf k}}^{(1)}=0.
\label{517}
\end{equation}
Comparing equation (\ref{517}) with the static equation for
$\phi_{\bf k}$ (equation \ref{316}), or the approximate form in the
inertial range (equation \ref{322}) that yields the leading order
behaviour that reads in terms of $\Lambda^{(1)}$ and
$\Lambda^{(2)}$
\begin{equation}
\int_{\infty}^{\infty}\Lambda^{(1)}(j,k)\phi_{\bf j}\phi_{\bf k}-\int_{\infty}^{\infty}\Lambda^{(2)}(j,k)\phi_{\bf j}\phi_{\bf{k}-\bf{j}}=0,
\label{518}
\end{equation}
we find that to leading order in $k$
\begin{equation}
\psi_{\bf k}=C\phi_{\bf k}, \label{519}
\end{equation}
where
\begin{equation}
C=\frac{\int_{-\infty}^{\infty}\Lambda^{(2)}\phi_{\bf j}\phi_{\bf{k}-\bf{j}}\;{\rm
d}j}{\int_0^k\Lambda^{(2)}(j,k)\phi_{\bf j}\phi_{\bf{k}-\bf{j}}\;{\rm d}j}
\label{520}
\end{equation}
can be shown to be independent of $k$. The faster decaying
contributions to the $\Lambda^{(2)}$ integral, coming from $j$'s
outside the segment $(0,k)$ will be matched against the faster
decays present in $\Phi_{\bf k}(t)$ (eq. \ref{55}). In more than one
dimension the situation is a bit more subtle. The reason is that
the long time decay of $\Phi_{\bf k}(t)$, $\exp({-\gamma kt^{1/\mu}})$,
can be matched against the slowest decay in the $\Lambda^{(2)}$
integral exactly as in the one dimensional case but that slowest
decay is contributed by a set of $\bf{j}$'s that is of measure
zero, i.e. $\bf{j}=\alpha\bf{k}$ with $0\leq\alpha\leq 1$. To
obtain the correct behaviour we need to consider not only the set
of $\bf{j}$'s where the decays can be matched exactly but also
the nearby vicinity
\begin{equation}
|\bf{j}|+|\bf{k}-\bf{j}|-k<B^{-1/\mu}(\gamma
t^{1/\mu})^{-1}=\epsilon, \label{521}
\end{equation}
where $\epsilon$ is small compared to $k$ (this is the basic long
time condition $\omega_{\bf k}t>>1$), such that the difference in the
decays is quite small.(The constant B is given by
$\omega_{\bf k}=Bk^{\mu}$). It is interesting to note that the effective
region of integration described by eq. (\ref{521}) is the interior
of an ellipsoid of revolution
\begin{equation}
\frac{(j_{\parallel}-k/2)^2}{a^2}+\frac{j_{\perp}^2}{b^2}=1,
\label{522}
\end{equation}
where $j_{\parallel}$ is the component of $\bf{j}$ in the
direction of $\bf{k}$, and $j_{\perp}$ is the part of $\bf{j}$
perpendicular to $\bf{k}$,
\begin{equation}
a=\frac{\epsilon +k}{2}\;\mbox{and}\;b=\sqrt{\frac{\epsilon
(\epsilon +2k)}{2}}. \label{523}
\end{equation}
Equation (\ref{512}) will be solved now by writing
\begin{equation}
\Phi_{\bf k}(t)=C\phi_{\bf k}(\omega_{\bf k}t)^{\beta}{\rm e}^{-\gamma
(\omega_{\bf k}t)^{1/\mu}}, \label{524}
\end{equation}
where $C$ is a dimensionless constant and restricting the
$\Lambda^{(2)}$ integration to the effective region described
above. We find that the form given above for $\Phi_{\bf k}(t)$ is
adequate to leading order provided \cite{SE2000}
\begin{equation}
\beta=\frac{d-1}{2\mu}\quad \label{525}
\end{equation}
The constant $C$ can be obtained explicitly as in the one
dimensional case as a ratio of two dimensionless integrals but it
is not of much importance.

\textbf{b}. The treatment presented above relies heavily on the
assumption that $\mu>1$. This is generic yet the most important,
perhaps, in the class of systems we study here is the noise driven
Navier-Stokes, which has $\mu=2/3$. To treat that particular
system, we go back to equation (\ref{313}) and write
\begin{equation}
\Phi_{\bf k}(t)=\phi_{\bf k}{\rm e}^{-\Gamma_{\bf k}(t)} \label{526}
\end{equation}
to yield
\begin{multline}
-\frac{\partial \Gamma_{\bf k}}{\partial t}=-\nu_{\bf k}-\int\Lambda^{(1)}\phi_{\bf j}\;{\rm d}^3j +\\
\int\Lambda^{(2)}\frac{\phi_{\bf j}\phi_{\bf{k}-\bf{j}}}{\phi_{\bf k}}{\rm
e}^{-\Gamma_{\bf j}(t)-\Gamma_{\bf{k}-\bf{j}}(t)+\Gamma_{\bf k}(t)}\;{\rm d}^3j.
\label{527}
\end{multline}
We subtract the above from the static equation (\ref{41}) (divided
by $\phi_{\bf k}$) and taking into account that $_{{\bf k}0}$ is zero for
finite $k$, we obtain
\begin{equation}
\frac{\partial \Gamma_{\bf k}}{\partial
t}=\int\Lambda^{(2)}\frac{\phi_{\bf j}\phi_{\bf{k}-\bf{j}}}{\phi_{\bf k}}\left(1-{\rm
e}^{-\Gamma_{\bf j}(t)-\Gamma_{\bf{k}-\bf{j}}(t)+\Gamma_{\bf k}(t)}\right)\;{\rm d}^3j.
\label{528}
\end{equation}
We assume now there is a scaling solution,
$\Gamma_{\bf k}(t)=\Gamma(\omega_{\bf k}t)$, and since $\omega_{\bf k}$ and $\phi_{\bf k}$
are each a power of $k$, we can scale out $t$ by the
transformation $k\rightarrow (Bt)^{1/\mu}k$, whereupon we find
\begin{multline}
\frac{3}{2}k^{1/3}\frac{\partial \Gamma(k)}{\partial k}=\frac{A}{B^2}\int\frac{\Lambda^{(2)}(\bf{k},\bf{j})j^{-11/3}}{k^{2/3}}\left(\frac{|\bf{k}-\bf{j}|}{k}\right)^{-11/3}\times\\
\left(1-{\rm
e}^{-\Gamma(j)-\Gamma(|\bf{k}-\bf{j}|)+\Gamma(k)}\right)\;{\rm
d}^3j. \label{529}
\end{multline}
Note that $A/B^2$ is dimensionless.

We consider now (\ref{529}) in the regime of very large $k$, and
assume that $\Gamma (k)\propto k^{\nu}$, with perhaps some
logarithmic correction.

We now split the integral on the right hand side of (\ref{529})
into contributions coming from regions where $j$ (or
$|\bf{k}-\bf{j}|$) is small compared to 1 and the region where
both are large compared to 1. Consider first the region where both
$j$ and $|\bf{k}-\bf{j}|$ are large. In that region, the term in
the exponent on the right hand side of (\ref{529}) is proportional
to $-[j^{\nu}+|\bf{k}-\bf{j}|^{\nu}-k^{\nu}]$. The discussion of
the first part of this section applies and the conclusion is that
\begin{equation}
\nu \,\leq \, 1.
\label{530}
\end{equation}
Consider next the region where $j$ is small compared to 1. Here we
cannot say that $\Gamma (j)$ is proportional to $j^{\nu}$. In
fact, it may start off as $j^{\alpha}$ for small $j$, and cross
over to $j^{\nu}$ for $j$'s of the order of 1. In any case,
consideration of the small $j$ dependence of the integrand in
equation (\ref{529}) reveals that it is enough to have $\alpha >
2/3$ in order to remove the convergence difficulties that may be
caused by the $j^{-11/3}$ factor. Now for $j$'s that are less that
1 we can approximate
\begin{equation}
\Gamma (|{\bf k}-{\bf j}|)\,-\,\Gamma({\bf k})\,=\,\left\{C\frac{\bf{j}\cdot\bf{k}}{k^2}+D\frac{j^2}{k^2}+E\frac{(\bf{j}\cdot\bf{k})^2}{k^4}\right\}k^{\nu},
\label{531}
\end{equation}
where the constants $C$, $D$ and $E$ are obtained from the
expansion. We will assume now that $\nu$ is actually less that 1
in order to study the consequences of that assumption. In the case
that $\nu <1$, the approximation presented by eq.(\ref{531}) is
indeed justified, because the term on the right hand side is small
due to the fact that $k>>1$. Note that this is not true for
$\nu=1$ where for $j$ of order 1 we are left on the right hand
side of (\ref{531}) with an expression that is of order 1.

The conclusion of all this is that if $\nu <1$ then for large $k$
\begin{equation}
\Gamma (j)+\Gamma(|\bf{k}-\bf{j})-\Gamma(k)>0,
\label{532}
\end{equation}
apart, perhaps, from a very small region of $j$'s near the origin
depending on the actual value of $\alpha$ and is smaller than
$k^{\nu -1}$, i.e. vanishingly small as $k$ becomes large, and
therefore of no consequence to the integral. The inequality holds
for large $j$ and large $|\bf{k}-\bf{j}|$ trivially, because
$\nu <1$ and for $j$ smaller than 1, the left hand side of
(\ref{532}) may, in fact, be approximated by $\Gamma(j)$. An
approximation that as stated above can fail, depending on the
value of $\alpha$, only for very small $j$'s such that
$j<k^{-\theta}$, with $\theta
>0$. The conclusion now is that the integrand in eq.(\ref{529}) is
positive definite and therefore contributions from different
regions sum up with no cancellation, (note that it is positive
definite only for large $k$ and $\nu <1$). The leading order $k$
contribution from the small $j$ integration, which we denote by
$I_<(k)$, can be easily shown to scale like
\begin{equation}
I_<(k)\,\propto \, k^{2/3} \,.
\label{533}
\end{equation}
This follows since for small $\bf{j}$,
$\Lambda^{(2)}(\bf{k},\bf{j})$ is proportional to
$\frac{k^{2}}{k^{2/3}}$. Because contributions are summed up, it
is clear that the leading $k$ dependence, $I(k)$, obtained from
the full integration must be such that
\begin{equation}
I(k)>I_<(k) \label{534} \, .
\end{equation}
Consequently, since $I(k)~\propto ~k^{\nu-2/3}$, as can be easily
seen from the left side of (\ref{529}) we arrive at
\begin{equation}
\nu\,\geq \, 4/3  \, .
\label{535}
\end{equation}
that is a contradiction to our assumption that $\nu <1$. Thus the
assumption $\nu<1$ is not consistent and we are left with $\nu=1$
as the only possibility. Hence we have shown that, to within a
possible logarithmic correction
\begin{equation}
\Gamma(x)\propto |x|.
\label{536}
\end{equation}

\newpage

\section{}
\setcounter{equation}{0}
 We have noted that the expansion
parameter of the series is $\frac{M^2\phi}{\omega^2}k^d$ and in
order for the series to scale, $\omega$ is determined by this
relation. Thus in KPZ at any order of accuracy, the equation
governing the index is $Z(\Gamma)=0$ and $Z$ is calculated using
an $\omega$ whose power is determined by
\begin{equation}
M^2\phi k^d\omega^{-1}=0(1).
\label{81}
\end{equation}

The front factor $B$ in $\omega_{\bf k}=Bk^{\mu}$ does not enter the
$Z(\Gamma)$ of (\ref{63}) and this is crucial for the self
consistency of the theory. Of course we have calculated
numerically using $Z_2(\Gamma)$ and get, say, $\Gamma_2$. When we
go as in (\ref{67}) to
\begin{equation}
Z_2+Z_4=0,
\label{82}
\end{equation}
the value of $\Gamma$ will certainly change for $KPZ$ (but
apparently not for KNS, for KNS is deficient in dimensions). One
can only check that $Z_4$ makes little difference by actually
calculating it, which is not at all easy from sheer algebraic
size, though we repeat the comment earlier that the numerical
value from $Z_2$ agrees excellently with numerical experiment.

It follows from this discussion that the only significance of an
$\omega_{\bf k}$ equation lies in the ability to calculate a front
factor. The series is formally valid for $any$ $\omega$, but we
have one powerful constraint in the value of a scaling relation.
We have chosen to use a time decay argument to derive (\ref{319})
but other methods are present in the literature. We could study
the Hermite operator and looked at, in harmonic oscillator
language, the first excited state above $P_0$ {\it {i.e.}} $hP_0$
and try to fit that \cite{Leslie}. However, all the equations used
have Galilean invariance,
\begin{equation}
h_{\bf k}\rightarrow h_{\bf k}{\rm e}^{i{\bf k} \cdot {\bf vt}}
\label{83}
\end{equation}
and the 'excited state' equations do not seem to have it, i.e. a denominator
\begin{equation}
\omega_{\bf k}+\omega_{\bf j}+\omega_{-k-j}\rightarrow\omega_{\bf k}+\omega_{\bf j}+\omega_{-k-j}+iv\cdot(\bf{k}+\bf{j}-\bf{k}-\bf{j})
\label{84}
\end{equation}
i.e. $\omega_{\bf k}+\omega_{\bf j}+\omega_{-k-j}$ is Galilean invariant; but
$\omega_{\bf j}+\omega_{-k-j}-\omega_{\bf k}$ is not, and it is this
combination that appears in the 'excited state' equations; they
also do not always converge. A difficulty of the same nature
arises in ref \cite{bouch} and if one uses the method of Herring
\cite{Herr} where the combinations $\omega_{\bf j}+\omega_{-k-j}$ appear
in the determination of static properties, that are not affected
by Galilean transformations. Another method is to argue that since
the equations should be valid for any $\omega$ one can minimise
the effect of variation of $\omega$ and differentiate (\ref{114})
w.r.t. $\omega_{\bf k}$. Yet another method is to maximise the "entropy"
in the $[h(r,t)]$ space \cite{McComb}. All these remarks apply
equally to $\Omega$ and, whereas for $\omega_{\bf k}$ one is only
searching for a front factor equation, with $\Omega_{{\bf
k}\omega}$ the whole complex plane presents itself. We feel that a
major problem has been uncovered here, and although we have
presented a practical solution to the problems, there is still a
great deal yet to be uncovered.

\newpage
\begin{figure}
\begin{center}
\includegraphics[width=12cm]{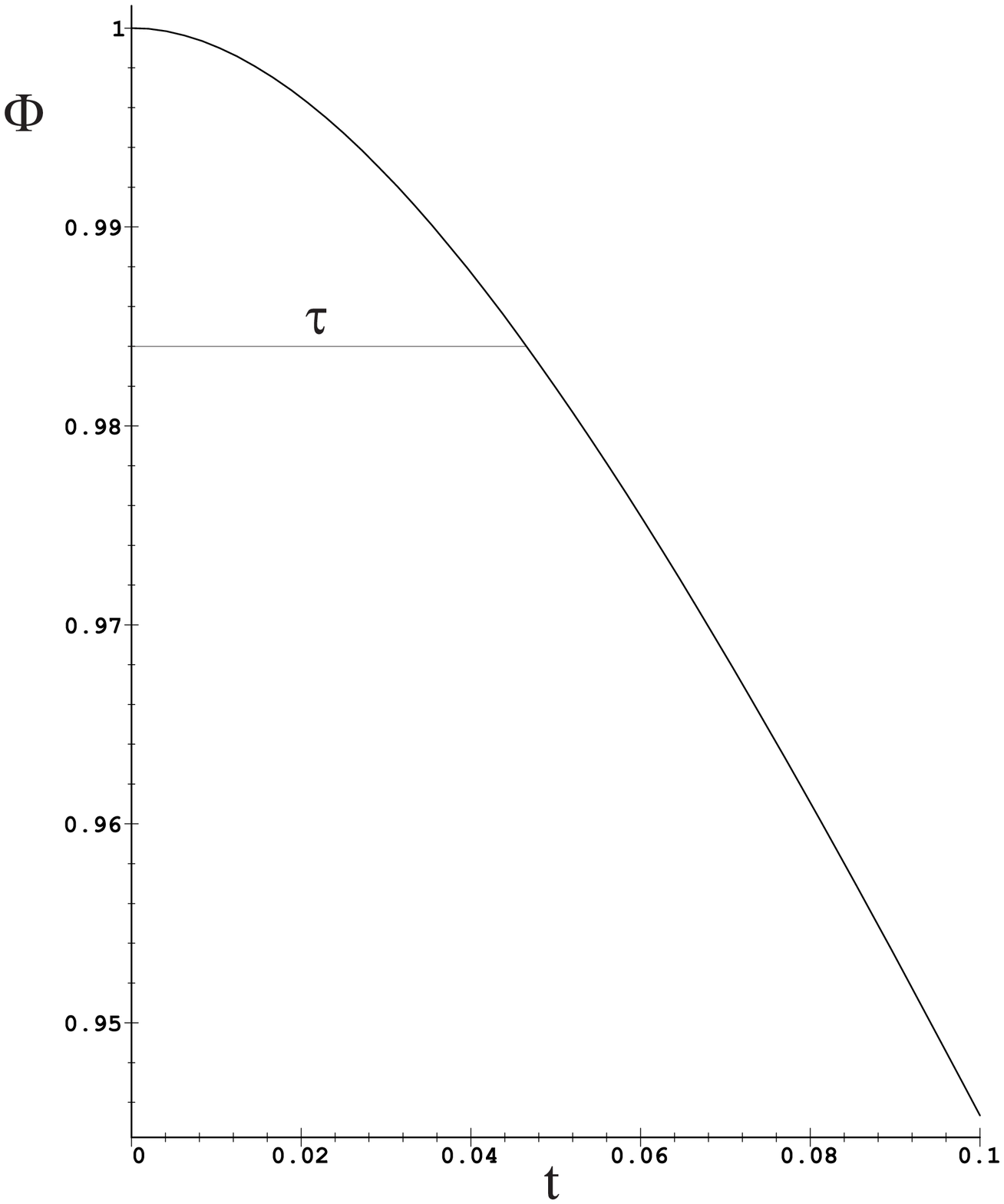}

\caption{The time dependent correlation $\Phi
  \left(t\right)=const\left[ \frac{e^{-\nu t}}{\nu}-
    \frac{e^{-bt}}{b}\right]$ for the ratio $\frac{b}{\nu} = 20$.}
\end{center}
\end{figure}

\newpage
\begin{figure}
\begin{center}
\includegraphics[width=12cm]{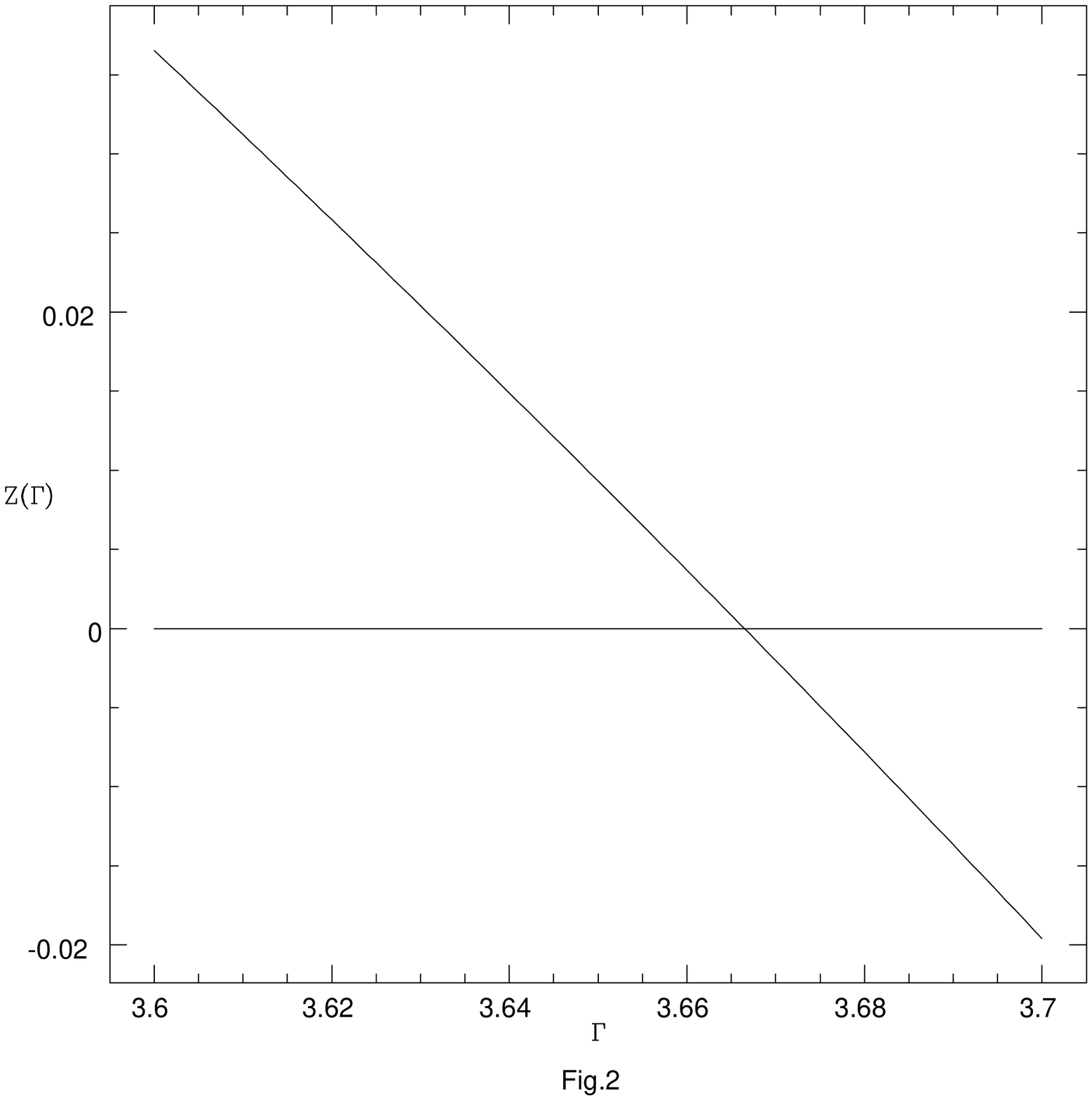}

\caption{The coefficient Z as a function of $\Gamma$ for the case of turbulence.}
\end{center}
\end{figure}

\end{document}